\documentclass[usenatbib]{mnras}
\usepackage{epsfig}
\usepackage{times}
\usepackage{amsmath,amssymb}
\usepackage{xcolor}
\usepackage{tikz}
\graphicspath{{figures/}}

  \def\simge{\mathrel{\raise1.16pt\hbox{$>$}\kern-7.0pt
    \lower3.06pt\hbox{{$\scriptstyle \sim$}}}}           
  \def\simle{\mathrel{\raise1.16pt\hbox{$<$}\kern-7.0pt
    \lower3.06pt\hbox{{$\scriptstyle \sim$}}}}           

\newcommand{\appropto}{\mathrel{\vcenter{
  \offinterlineskip\halign{\hfil$##$\cr
    \propto\cr\noalign{\kern2pt}\sim\cr\noalign{\kern-2pt}}}}}

\title[Hot pre-white dwarfs from SALT]{Hot white dwarfs and pre-white dwarfs discovered with SALT\footnote{*based on observations made with the Southern African Large Telescope (SALT) and at the the South African Astronomical Observatory}}
    
\author[C. S. Jeffery et al.]{C. S. Jeffery $^1$\thanks{email: simon.jeffery@armagh.ac.uk}, 
K. Werner $^2$, 
D. Kilkenny $^3$, 
B. Miszalski $^{4}$, 
I. Monageng $^{5,6}$, 
and E. J. Snowdon $^1$ \\
  1. Armagh Observatory and Planetarium, College Hill, Armagh BT61 9DG, United Kingdom\\
  2. Institut f\"ur Astronomie und Astrophysik, Kepler Center for Astro and Particle Physics, Universit\"at T\"ubingen, Sand 1, 72076 T\"ubingen, Germany\\
  3. Department of Physics \& Astronomy, University of the Western Cape, Private Bag X17, Belville 7535, South Africa \\
  4. Australian Astronomical Optics, Faculty of Science and Engineering, Macquarie University, North Ryde, NSW 2113, Australia\\ 
  5. Department of Astronomy, University of Cape Town, Private Bag X3, Rondebosch 7701, South Africa\\
  6. South African Astronomical Observatory, PO Box 9, Observatory 7935, Cape Town, South Africa\\
  }

\date{Accepted 2022 November 29; Received 2022 November 14; in original form 2022 October 17.}
\pagerange{\pageref{firstpage}--\pageref{lastpage}} 
\pubyear{2023} 

\begin{document}
\label{firstpage}
\maketitle

\begin{abstract}
 The Southern African Large Telescope (SALT) survey of helium-rich hot subdwarfs aims to explore evolutionary pathways amongst groups of highly-evolved stars. The selection criteria mean that several hot white dwarfs and related objects have also been included. This paper reports the discovery and analysis of eight new very hot white dwarf and pre-white dwarf stars with effective temperatures exceeding 100\,000\,K. They include two PG1159 stars, one DO white dwarf, three O(He) and two O(H) stars. One of the O(H) stars is the central star of a newly-discovered planetary nebula, the other is the hottest `naked' O(H) star.  Both of the PG1159 stars are  GW\,Vir variables, one being the hottest GW\,Vir star measured and a crucial test for pulsation stability models. The DO white dwarf is also the hottest in its class.  
\end{abstract}

\begin{keywords}
surveys, white dwarfs, stars: early-type, stars: fundamental parameters, stars: oscillations, planetary nebulae: individual
\end{keywords}

\section{Introduction}

This paper describes how a survey of helium-rich subdwarf stars (He-sds) has led to the discovery of several very hot white dwarf and pre-white dwarf stars. 
He-sds cover a range of effective temperature from around 30\,000 K to well over 60\,000 K and comprise some 10\% of all hot subdwarfs. 
They are considered to form a population distinct from the hydrogen-rich subdwarfs, having clearly different evolution histories with the majority likely being the result of low-mass double white dwarf mergers \citep{zhang12a}. 
However, their properties are diverse and have led to a spectroscopic survey to characterise a large sample with which to explore evolutionary connections \citep{jeffery21a}.
The survey uses the Southern African Large Telescope (SALT) at both low-resolution ($\sim2.2$\AA) and, where possible,  high-resolution ($\sim0.1$\AA). 
Approximately 300 He-sds have now been observed and classified (Jeffery et al. in preparation).
The sample aims to include any star visible to SALT with a classification of He-sdB, He-sdOB or He-sdO and Gaia magnitude $G\simle16.5$.
Additional targets were selected by colour and absolute magnitude from the GAIA survey \citep{geier19}. 

A consequence of the selection criteria is that several white dwarfs have been included. 
The presence of central emission in the He{\sc ii} 4686\AA\ line prompted us to look more closely at some of these, with the expectation that they could be very hot white dwarfs, including PG1159 stars and DO white dwarfs.   
Both classes include massive white dwarfs with CO or ONeMg cores and, broadly speaking, are considered to lie on the cooling track from the the asymptotic giant branch to the white dwarf sequence, roughly at the point where the track transits from contraction at constant luminosity to cooling at constant radius. 
PG1159 stars are widely considered to be the result of a late or very-late thermal pulse in a post-AGB star or in a white dwarf, respectively, and also to be the precursors of the DO white dwarfs or DA white dwarfs, when they retain some hydrogen in their envelope \citep{werner06}. 
40 years after discovery \citep{mcgraw79}, the PG\,1159 stars now number approximately 60 \citep{werner06,reindl16,werner22} and include some of the hottest stars known (e.g. H1504+65, RX\,J0439.8$-$6809; \citet{werner15}). The number of known DO white dwarfs currently stands at around 160, but hot DOs (defined as those that do not exhibit neutral helium lines, i.e., with effective temperatures of 100\,000\,K and higher, \citep[e.g.][]{dreizler96} are quite rare \citep[about 15;][the Montreal White Dwarf Database\footnote{\tt https://www.montrealwhitedwarfdatabase.org}]{dufour17}. Related hot pre-white dwarfs are the O(H) and O(He) classes introduced by \citet{mendez91}, denoting H-rich and He-rich objects that are in most cases central stars of old planetary nebulae. 
This broad picture almost certainly obscures a more complex and diverse mixture of origins for all groups, indicated by a variety of surface chemistries, pulsation properties and associations with planetary nebulae.   
Continuing to build a census of all classes is essential for developing our picture of the late stages of stellar evolution. 

This paper presents the SALT survey observations of white dwarfs obtained to date (\S\,2). 
It includes an analysis of the new discoveries (\S\,3), an inspection of the data for  planetary nebulae (\S\,4), and a photometric study of both PG1159 stars in the sample (\S\,5). 
The conclusion (\S\,6) summarises the new discoveries in the wider context of very hot white dwarfs. 

\section{SALT/RSS Observations}

\begin{figure*}
\includegraphics[width=0.9\linewidth]{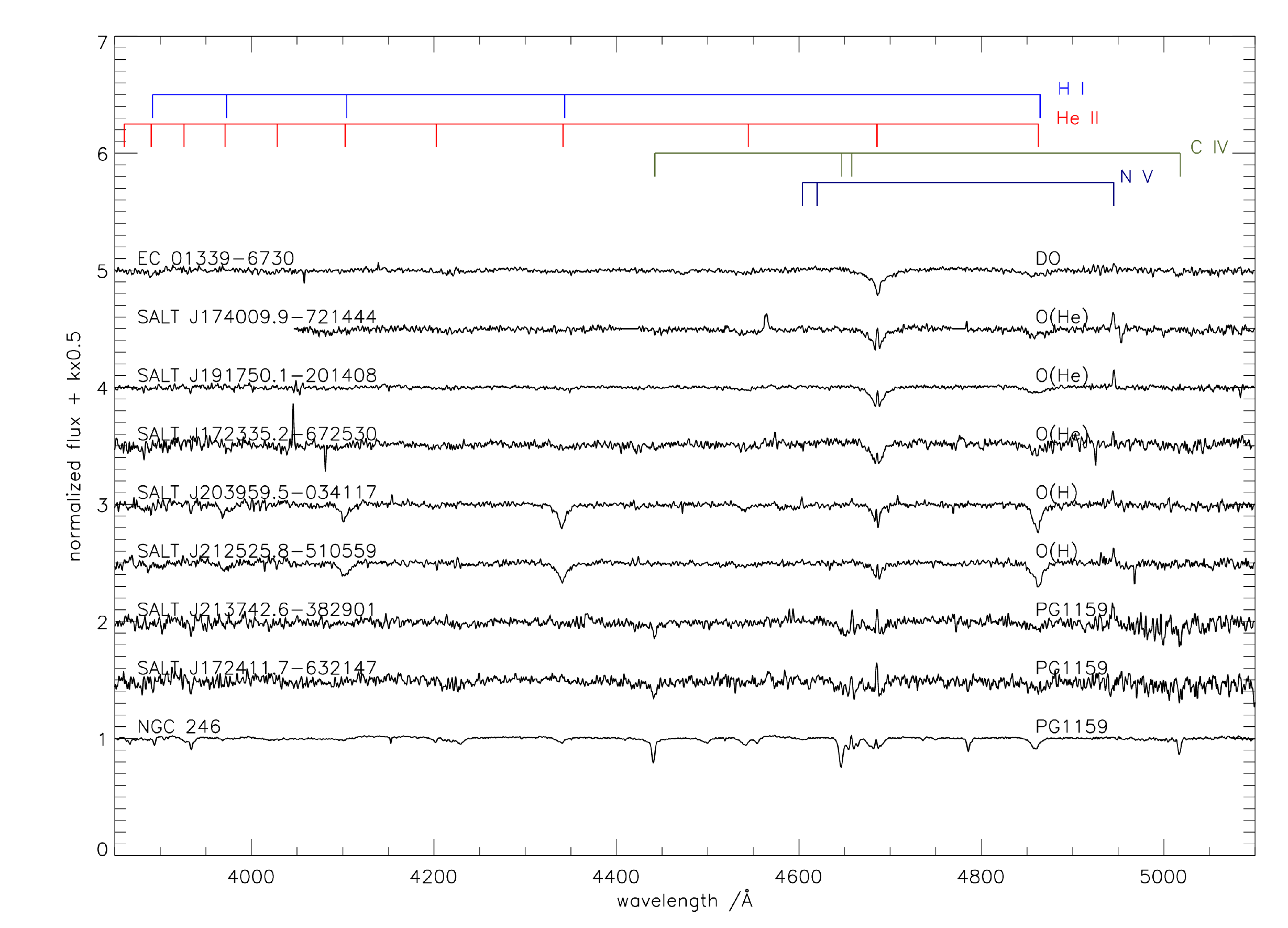}
\caption{SALT/RSS spectra of hot white dwarfs and pre-white dwarfs. Identifiers are on the left, spectral classes are on the right. Positions of principal lines are indicated at the top. The data are slightly smoothed (FWHM=0.5\AA). Gaps in the spectrum of J1740 are replaced by continuum.  The spectrum of a well-known PG1159 star, the central star of NGC\,246, is included for comparison. }
\label{f_rss1}
\end{figure*}

\begin{table*}
    \centering
    \begin{tabular}{llccll}
    \hline
    SALT identifier  &  Gaia DR3  & $G$ & $p$ (mas) &  SALT Observation (UTC) & Total exposure (s) \\
    \hline
    EC 01339$-$6730         & 4698459205509788288 & 16.77 & 2.12$\pm$0.05 & 20200623 04:00 & 1500  \\
    SALT J172335.2$-$672530 & 5811791728816787584 & 16.28 & 1.14$\pm$0.06 & 20190906 19:00 & 1000 \\
    SALT J172411.7$-$632147 & 5910236846008692352 & 16.59 & 0.59$\pm$0.06 & 20190908 18:45 & 1000 \\
    SALT J174009.9$-$721444 & 5803795977174249344 & 16.11 & 1.00$\pm$0.04 & 20200701 21:55 & 1000 \\
    SALT J191750.1$-$201408 & 4083008911902900992 & 15.97 & 1.20$\pm$0.06 & 20201005 19:30 & 1000 \\
    SALT J203959.5$-$034117 & 4224550035873178240 & 16.92 & 0.15$\pm$0.08 & 20200909 21:30 & 1000 \\
    SALT J212525.8$-$510559 & 6466250289796929024 & 16.82 & 0.34$\pm$0.07 & 20220625 23:00 & 1000 \\
    SALT J213742.6$-$382901 & 6585736932806500736 & 16.95 & 0.54$\pm$0.09 & 20200825 18:50 & 1000 \\
    \hline
    \end{tabular}
    \caption{The SALT hot white dwarf and pre-white dwarf sample showing: SALT or other catalogue identifiers, Gaia catalogue numbers, Gaia $G$ magnitudes and parallaxes \citep{gaia21.dr3}, SALT observation dates and times, and total spectroscopy exposure times. }
    \label{t_gaia}
\end{table*}

Commencing in 2018, observations have been obtained with the SALT Robert Stobie Spectrograph (RSS) at the Southern African Large Telescope (SALT). 
The observing method and data reduction procedures are described by \citet{jeffery21a}.
The spectrograph detector consists of three adjacent CCDs, each separated by a small gap. 
To avoid gaps in the observed spectrum, observations were obtained at two camera angles covering overlapping spectral ranges and then merged into a single data product. 
Occasionally, weather conditions or technical problems led to data from only one camera position being useable (cf. SALT J174009.9-721444: Fig. \ref{f_rss1}).  

Targets were initially selected from stars classified He-sdO, He-sdB, He-sdOB, sdOD, or their equivalents, in a wide range of catalogues \citep{jeffery21a}.
Additional targets were taken from the colour-selected list of hot subdwarf candidates provided by \citet{geier19}. 
Most of these additional targets only exist in the Gaia catalogue \citep{gaia21.dr3}. 
To assist with observing and data management, we have assigned a coordinate-based identifier to each target. Equivalences are given in Table\,\ref{t_gaia}. For convenience within this paper, individual stars may be abbreviated to Jhhmm, e.g. J1724 $\equiv$ SALT J172411.7$-$632147. 

For much of the campaign, these targets were observed at Priority 4, {\it i.e.} during gaps in the telescope queue. 
Latterly, additional time was awarded to accelerate the execution of survey.  
The extension of the survey to colour-selected candidates inevitably led to observations of an increasing fraction of stars from outside the original scope of the project, including those presented here. 
 
The resolution of 2.2\AA\ matches that used for spectral classification by \citet{drilling13}. Relative line strengths for specific hydrogen and helium lines provide preliminary classifications, which are then checked manually. White dwarfs are easily identified as outliers (Fig. \ref{f_rss1}).

The outliers include white dwarfs showing He{\sc i} 4471\AA, including seven DB, two DBA and one DAB white dwarfs. 
Being mostly taken from the Edinburgh-Cape survey of faint blue stars \citep{stobie97a,kilkenny97,odonoghue13,kilkenny15,kilkenny16}, the majority are previously known to science \citep{bergeron11,jeffery21a}. 
Eight additional outliers are clearly hotter, showing significant absorption and/or emission at He{\sc ii} 4686\AA. These are the subject of this paper. 

\subsection{Hot hydrogen-rich pre-white dwarfs}

Two stars exhibit Balmer lines plus He{\sc ii} 4686\AA\ with a central emission but no He{\sc i} lines, pointing at a very high effective temperature. They also have very weak emission from N{\sc v} 4604/4620 and 4945\AA\ and C{\sc iv} 4658\AA\ (Fig.\ref{f_rss1}). They can be classified as O(H) stars using the scheme of \citet{mendez91}.

\subsection{Hot helium-rich (pre-) white dwarfs}

Four stars display spectra dominated by He{\sc ii} lines while neutral He lines are absent, again pointing at very high effective temperatures. One of these (EC 01339$-$6730) is a hot DO white dwarf while, because of lower surface gravity (see below), the other three are  O(He) stars according to the \citet{mendez91} classification scheme. They show weak emission from N{\sc v} 4945\AA. The observation of SALT J174009.9$-$721444 failed for the second grating angle. 

\subsection{PG 1159 stars} One star included in the 2021 SALT sample  as a comparison for classification was BD$-12^{\circ}134$A, the central star of the planetary nebula NGC\,246 \citep{dreyer88} and also a PG1159 star. The second part of the SALT survey includes two new PG1159 stars (SALT J213742.6$-$382901 = Gaia EDR3 6585736932806500736 and SALT J172411.7$-$632147 = Gaia EDR3 5910236846008692352). Both were identified as hot subdwarfs from photometry by \citet{geier19}. The first was also identified as a white dwarf by \citet{gentilefusille19}. These hitherto unknown PG1159 stars were recognised by their resemblance with BD$-12^{\circ}134$A and show the hallmarks typical of this spectral class, an absorption trough comprising He{\sc ii} 4686\AA\ and C{\sc iv} 4647, 4658\AA, plus other weak C{\sc iv} lines. The first also exhibits emission lines from N{\sc v} 4604/4620 and 4945\AA.


\begin{figure*}
\begin{center}
\includegraphics[width=0.9\linewidth]{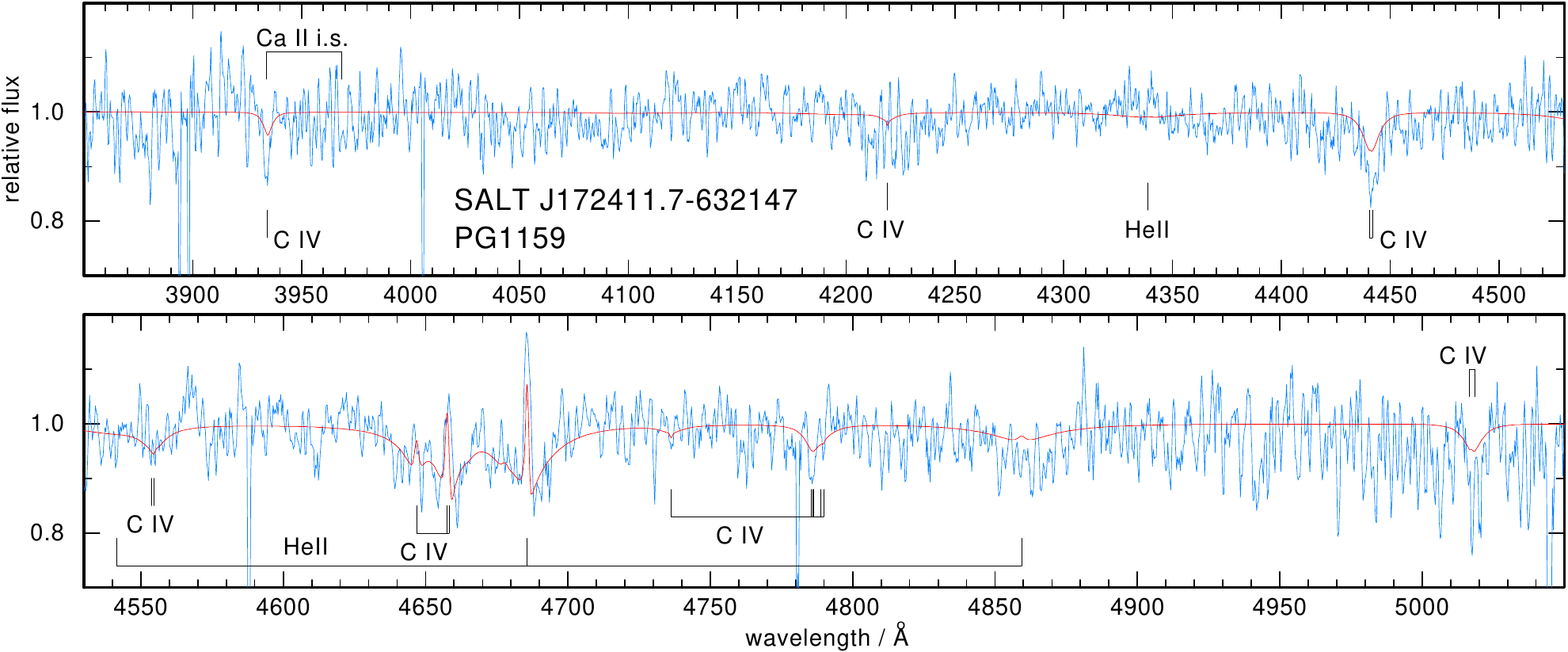}\vspace{3mm}
\includegraphics[width=0.9\linewidth]{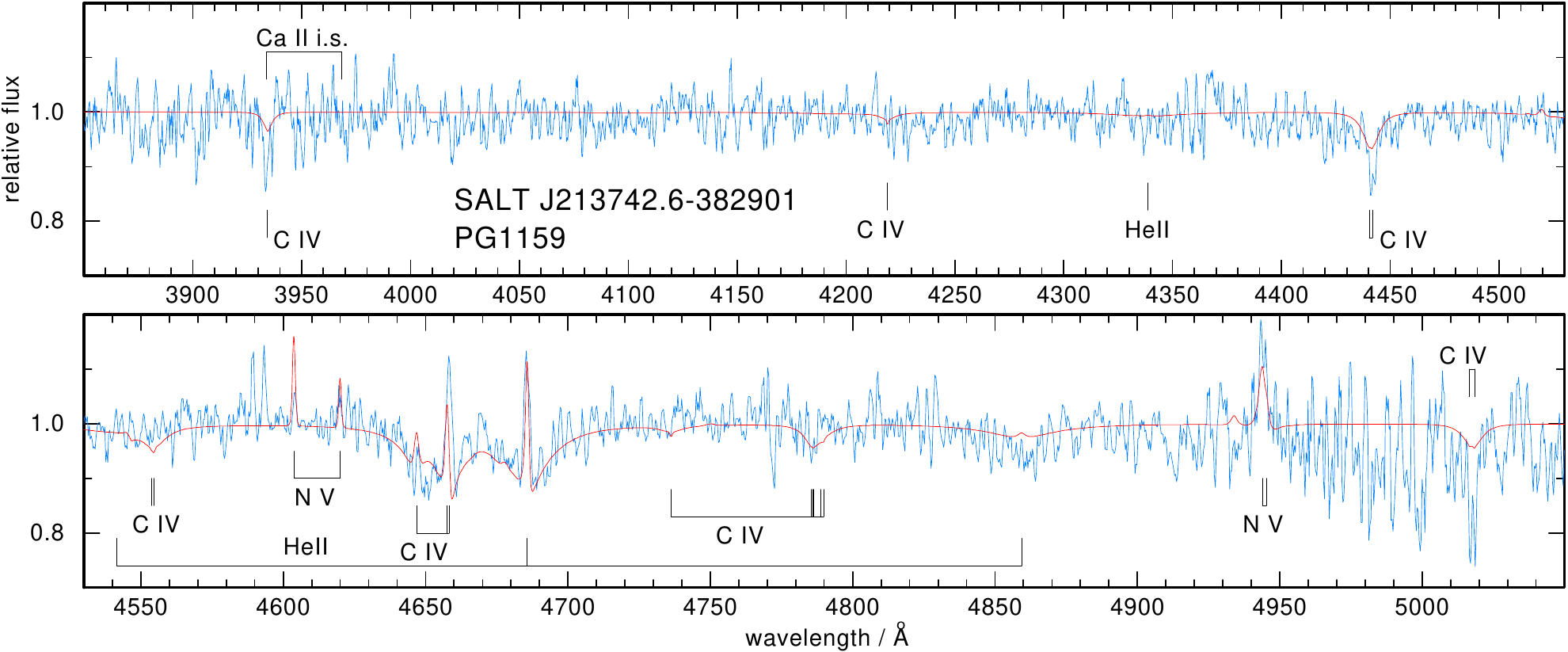}
\caption{Normalised SALT/RSS spectra (smoothed with a 0.5\AA\ wide boxcar) of two PG1159 stars as labelled (blue) with best fit non-LTE model spectra (red). Significant lines due to principal ions in the model are identified, as well as the location of the calcium interstellar lines. The model atmosphere parameters are shown in Table\,\ref{t_pars}.}
\label{f_fits1}
\end{center}
\end{figure*}

\begin{figure*}
\begin{center}
\includegraphics[width=0.9\linewidth]{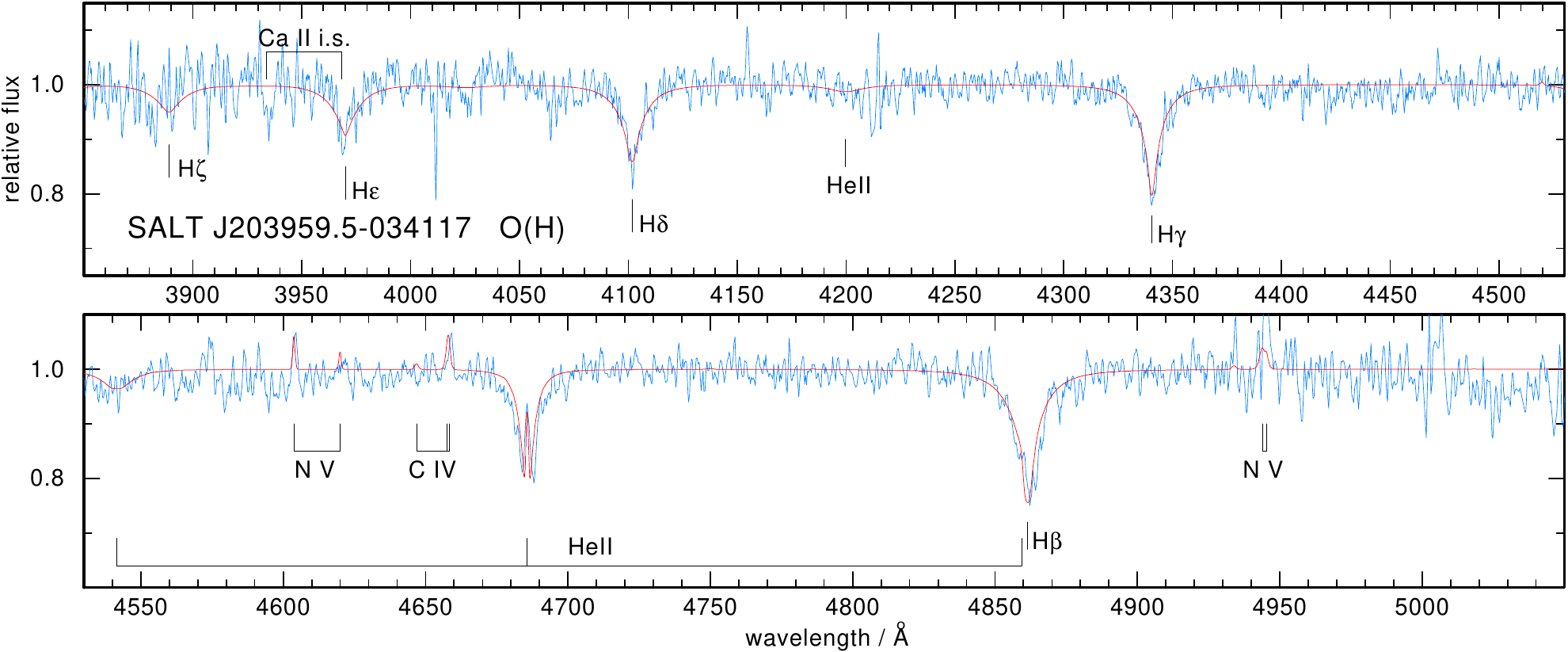}\vspace{3mm}
\includegraphics[width=0.9\linewidth]{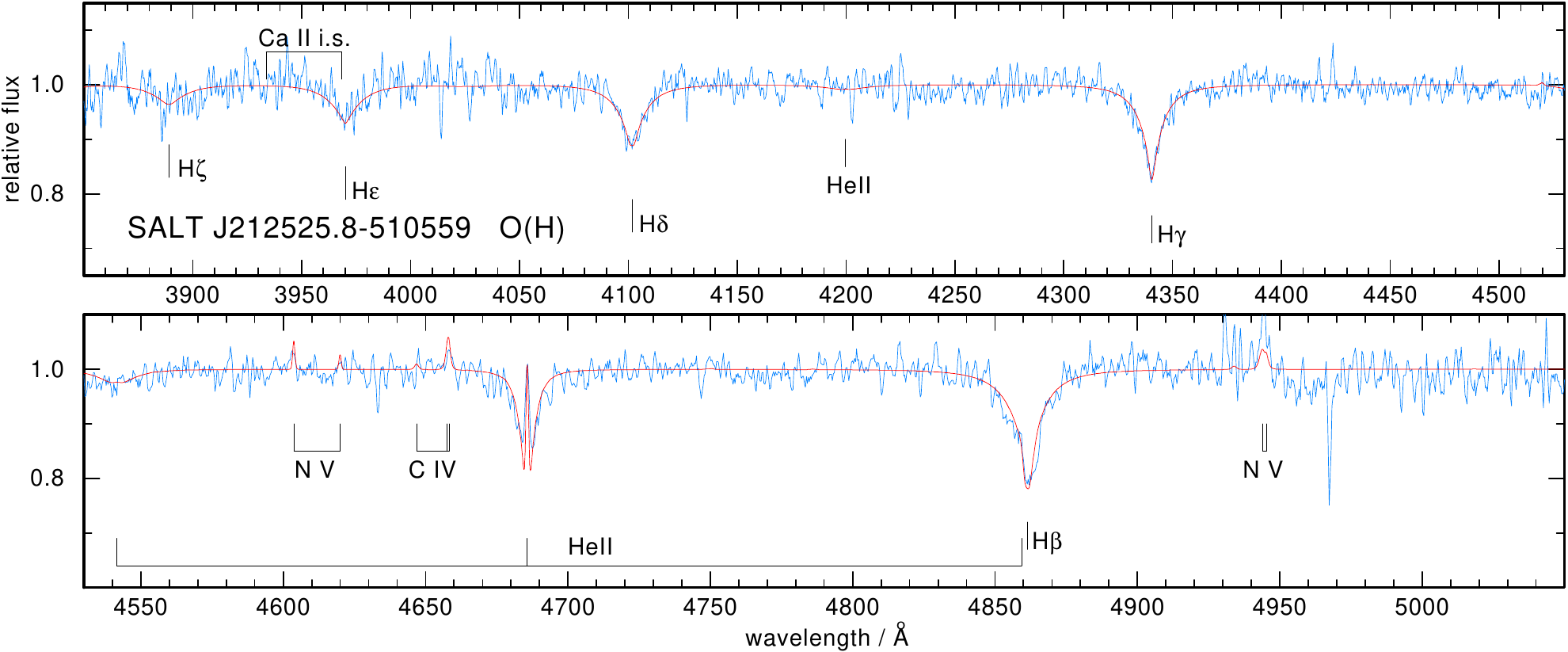}\vspace{3mm}
\includegraphics[width=0.9\linewidth]{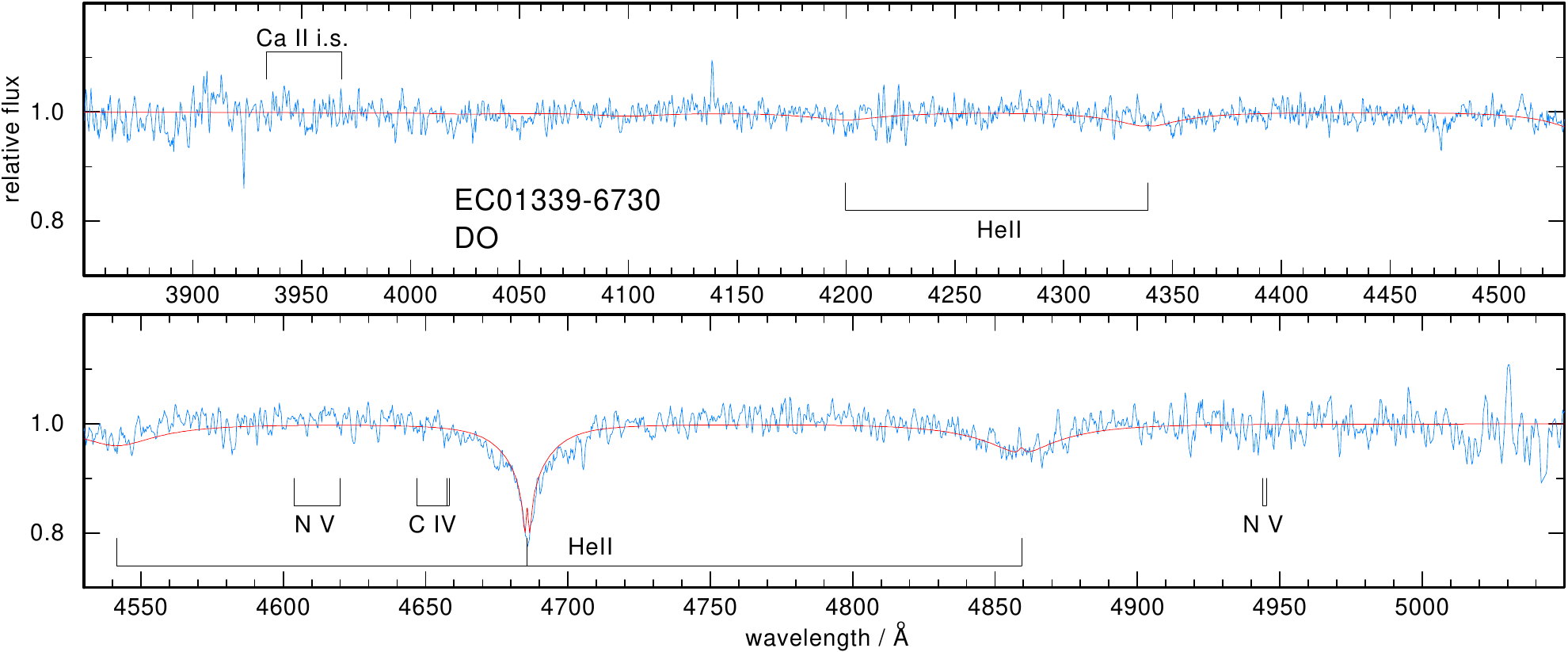}
\caption{Like Fig.\ref{f_fits1} but here for two O(H) stars and one DO white dwarf.}
\label{f_fits2}
\end{center}
\end{figure*}

\begin{figure*}
\begin{center}
\includegraphics[width=0.9\linewidth]{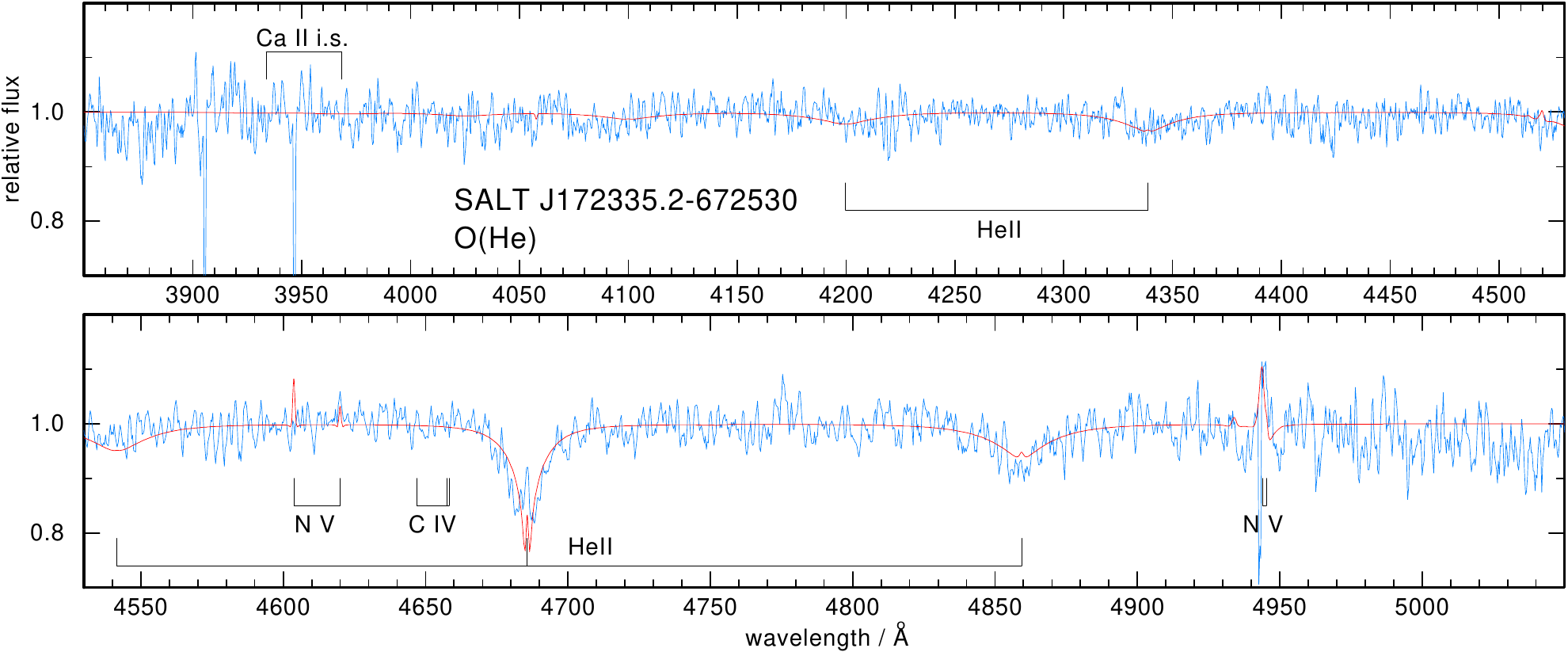}\vspace{3mm}
\includegraphics[width=0.9\linewidth]{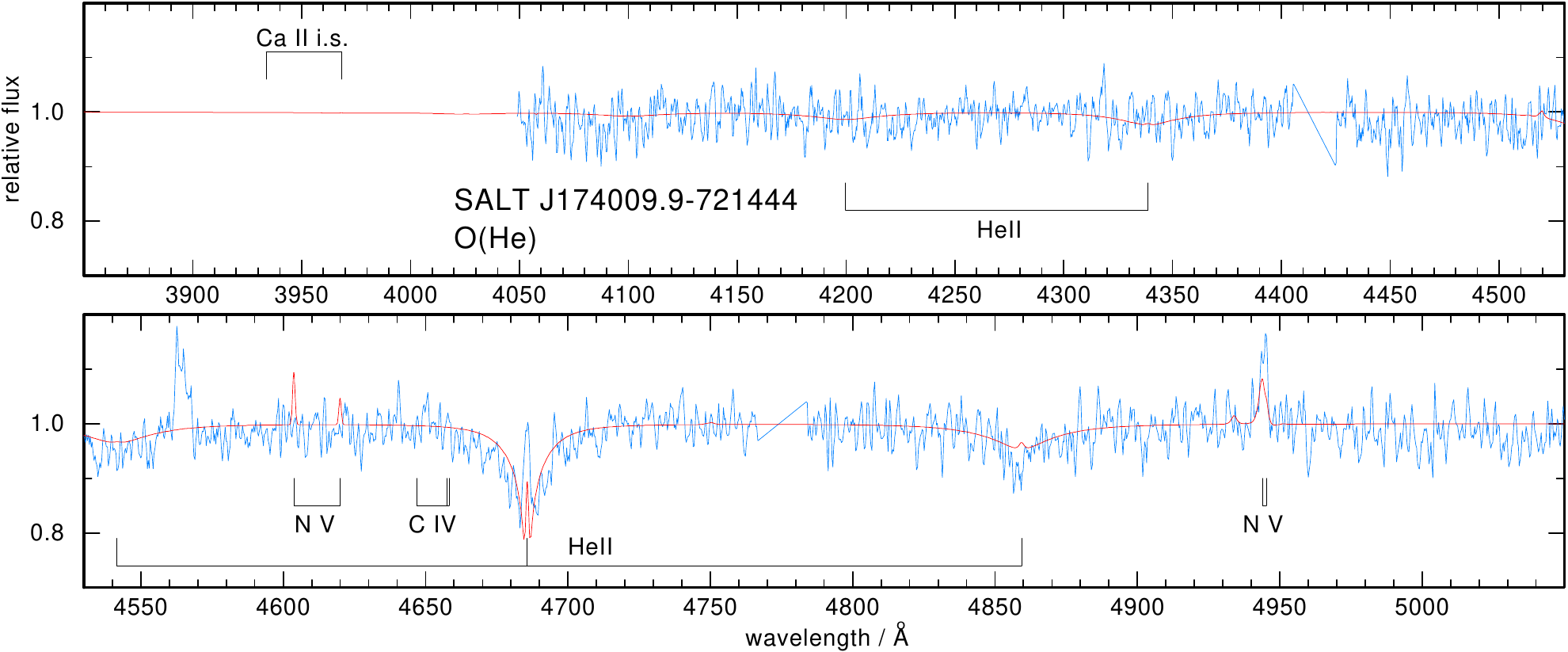}\vspace{3mm}
\includegraphics[width=0.9\linewidth]{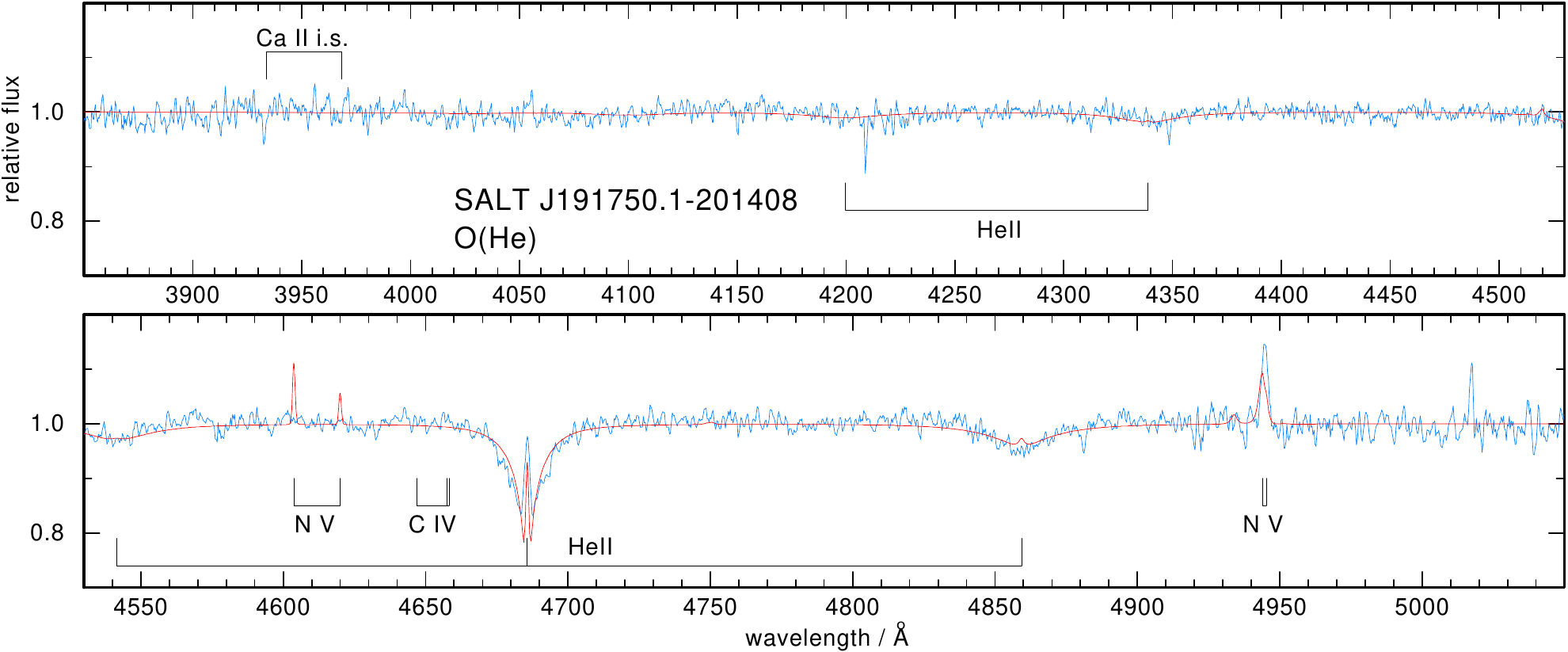}\vspace{3mm}
\caption{
As Fig.\ref{f_fits1} but here for three O(He) stars.
}
\label{f_fits3}
\end{center}
\end{figure*}

\begin{figure*}
\begin{center}
\includegraphics[width=0.8\linewidth]{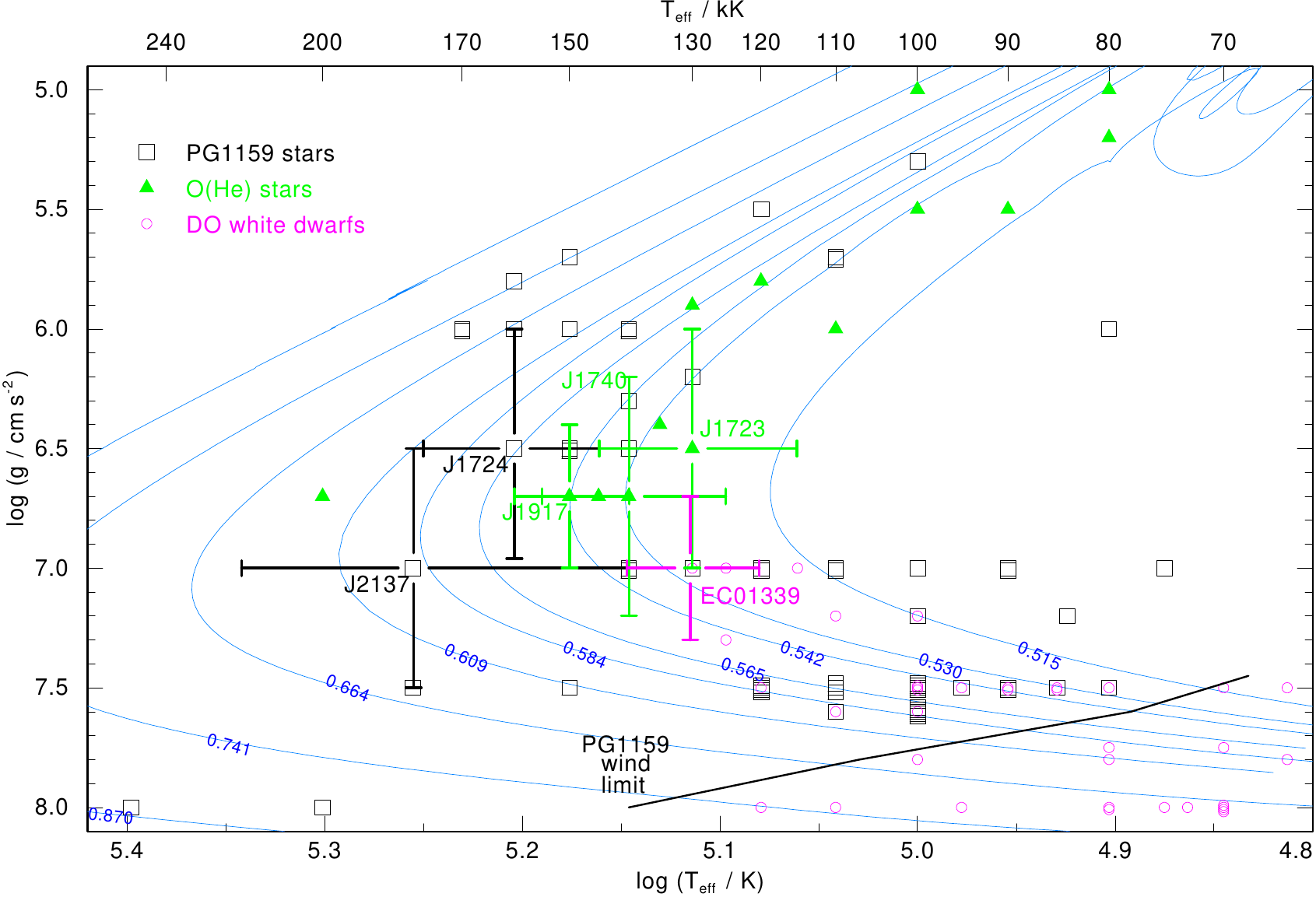}
\caption{Positions and error bars of the new DO white dwarf, PG1159 and O(He) stars in the Kiel diagram together with known objects of these classes. Evolutionary tracks by \citet{althaus09} are labelled with the stellar mass in solar units. }
\label{fig_gteff}
\end{center}
\end{figure*}

\begin{figure}
\begin{center}
\includegraphics[width=\linewidth]{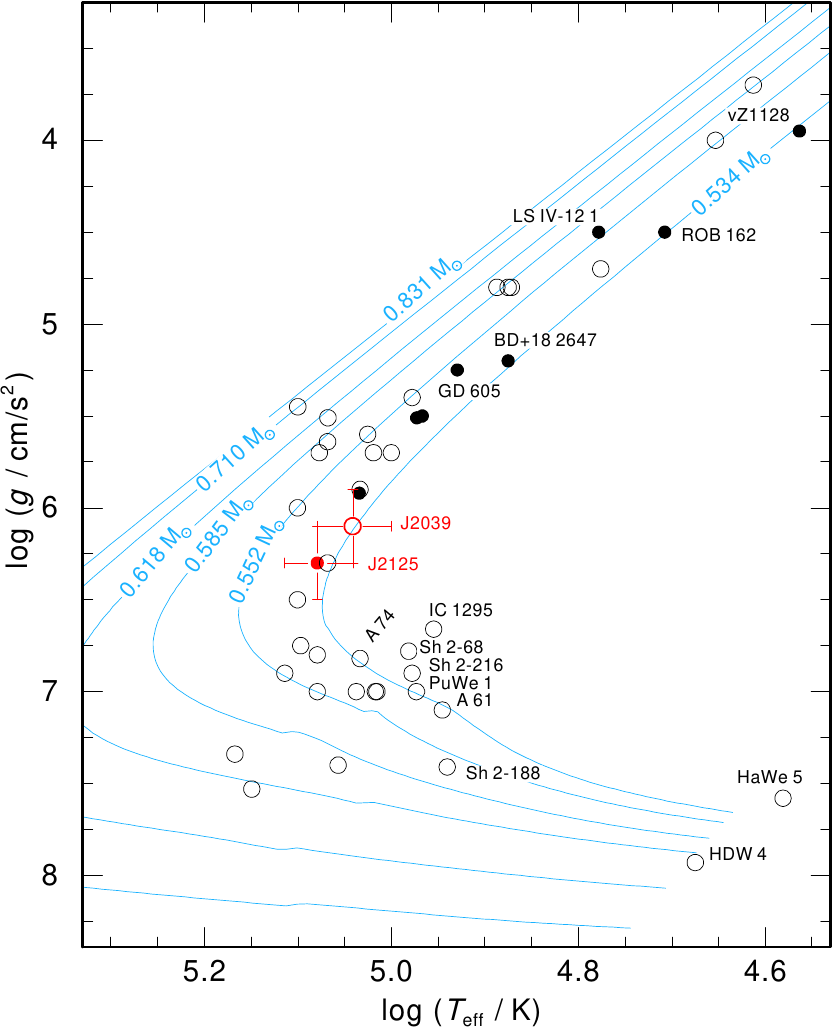}
\caption{Positions of the two new O(H) stars (red symbols) in the Kiel diagram together with other objects of this class. Evolutionary tracks by \citet{millerbertolami16} are labelled with the stellar mass in solar units. H-rich central stars of PNe are depicted by open circles and naked O(H)-type stars by filled circles. This is an updated plot from \citet{reindl16}; see references in the caption of their Fig.\,3.}
\label{fig_gteff_oh}
\end{center}
\end{figure}

\begin{table*}
    \centering
    \begin{tabular}{llcccccc}
    \hline
    Star                  & Class & $T_{\rm eff}$/K     & $\log g/{\rm cm\, s^{-2}}$   &  H            & He            & C             & N  \\
    \hline
    EC 01339-6730         & DO     & 130\,000$\pm$10\,000 & 7.0$\pm$0.3 & $-$           & 1.0           & $-$           & $-$ \\[2mm]
    SALT J174009.9$-$721444 & O(He)  & 140\,000$\pm$15\,000 & 6.7$\pm$0.5 & $-$           & 0.997         & $-$           & 0.003$\pm$0.5\,dex\\
    SALT J191750.1$-$201408 & O(He)  & 150\,000$\pm$10\,000 & 6.7$\pm$0.3 & $-$           & 0.997         & $-$           & 0.003$\pm$0.5\,dex\\
    SALT J172335.2$-$672530 & O(He)  & 130\,000$\pm$15\,000 & 6.5$\pm$0.5 & $-$           & 0.99          & $-$           &  0.01$\pm$0.5\,dex\\[2mm]
    SALT J203959.5$-$034117 & O(H)   & 110\,000$\pm$10\,000 & 6.1$\pm$0.2 & 0.75$\pm$0.05 & 0.25$\pm$0.05 & $8\cdot10^{-4}\pm$0.5\,dex   & $2.3\cdot10^{-4}\pm$0.5\,dex\\
    SALT J212525.8$-$510559 & O(H)   & 120\,000$\pm$10\,000 & 6.3$\pm$0.2 & 0.75$\pm$0.05 & 0.25$\pm$0.05 & $8\cdot10^{-4}\pm$0.5\,dex   & $2.3\cdot10^{-4}\pm$0.5\,dex\\[2mm]
    SALT J172411.7$-$632147 & PG1159 & 160\,000$\pm$20\,000 & 6.5$\pm$0.5 & $-$           & 0.50$\pm$0.25 & 0.50$\pm$0.25 & $-$        \\
    SALT J213742.6$-$382901 & PG1159 & 180\,000$\pm$40\,000 & 7.0$\pm$0.5 & $-$           & 0.49$\pm$0.25 & 0.49$\pm$0.25 & 0.01$\pm$0.3\,dex \\
    \hline
    \end{tabular}
    \caption{Results from non-LTE model atmosphere fits for hot (pre-) white dwarfs discovered with SALT. Element abundances are given in mass fractions.}
    \label{t_pars}
\end{table*}

\section{Spectral Analysis}

We used the T\"ubingen Model-Atmosphere Package (TMAP) to build grids of non-LTE plane-parallel models in radiative and hydrostatic equilibrium. For the two PG1159 stars we used models of the type introduced by \citet{werner14}. In essence, they include H, He, C, and O. Since H and O are not detected in the observed spectra, we set the respective model abundances to very low values such that they do not affect the spectra in the observed wavelength region. One of the PG1159 stars exhibits N{\sc v} emission lines at 4604/4620\,\AA\ and at 4945\,\AA. In the respective models non-LTE line-formation iterations were performed for nitrogen. Best-fit models within this grid were identified to provide effective temperatures, surface gravities and element abundances. The resulting fits are shown in Fig.~\ref{f_fits1} and the model parameters are shown in Table~\ref{t_pars}. Errors in atmospheric parameters were estimated by comparing synthetic spectra from our grid with the observations. Both PG1159 stars turn out to be among the hottest in their class, albeit the error for the effective temperature of SALT J213742.6-382901 is particularly large (180\,000$\pm$40\,000\,K). Observations in a broader optical wavelength range would reveal more temperature sensitive lines of C{\sc iv} and O{\sc vi} that would yield stronger constraints. 

For the three O(He) stars and the DO white dwarf we computed a grid of pure helium models and, as for the PG1159 stars, nitrogen line formation was performed. Again, best-fit models were chosen by eye (Figs.~\ref{f_fits2} and \ref{f_fits3}). We find that the four He-rich stars are very hot, having temperatures in the range 130\,000--150\,000\,K and $\log g / {\rm cm s^{-2}}$ around 7 (Table~\ref{t_pars}). The three O(He) stars exhibit a N{\sc v} 4945\,\AA\ emission feature but not the 4604/4620\,\AA\ doublet, while in the models with the adopted atmospheric parameters the 4604/4620\,\AA\ doublet is seen in emission. The models show that at slightly lower temperatures of around 120\,000\,K, the 4604/4620\,\AA\ lines turn from emission to absorption, being invisible at $\sim$120\,000\,K, while 4602/4620\,\AA\ is still in emission unless the N abundance is reduced from N=0.01 to N=0.001. It could be that this emission/absorption turning point would appear in hotter models when more realistic temperature structures are considered. The upper atmospheric layers could be affected by trace metals which are not included in the current models. As a consequence, we adopted a rather large error for the N abundance (0.5\,dex). 

The analysis of the two O(H) stars proceeded in a similar way. We computed models comprising hydrogen and helium to determine effective temperature, surface gravity, and H/He abundance ratio. Additional line-formation calculations were performed to find C and N abundances (Table~\ref{t_pars}, Fig.~\ref{f_fits2}). Both objects have temperatures just above 100\,000\,K and relatively low surface gravities of around $\log g\sim 6$. Their H/He ratio is solar. The C and N abundances are 1/3 solar but within the error limits they are compatible with solar abundances.

The positions of the analysed PG1159 stars, O(He) stars, and the hot DO white dwarf in the Kiel ($\log g - T_{\rm eff}$) diagram are shown in Fig.\ref{fig_gteff}, together with other objects of these classes. The two new PG1159 stars are remarkable because they belong to the hottest members of their group \citep{werner06} that currently comprises about 60 stars with effective temperatures between 75\,000 and 250\,000\,K and gravities between $\log g = 5.3$ and 8. O(He) stars are quite rare \citep{reindl14}  and the three new members extend the group to 14 stars covering 80\,000--200\,000\,K and $\log g = 5.0-6.7$. The new O(He) stars are at the high-temperature and high-gravity end of the O(He) regime. To our best knowledge, the new hot DO is the hottest DO white dwarf known (130\,000\,K), rivalling the hitherto hottest one, PG0038+199 (125\,000\,K) \citep{werner17}. 

The positions of the two O(H) stars in the Kiel diagram are shown in Fig.\,\ref{fig_gteff_oh}, together with other O(H)-type stars with and without a planetary nebula. The unexpected phenomenon of the ``missing'' PN for the O(H) stars is known for just a few objects and has been discussed e.g. by \citet{reindl16}. Accordingly, ``naked'' stars are more common among the PG1159 and O(He) classes and can be explained by a (very) late thermal pulse evolution and binary WD mergers, respectively. 


\begin{figure}
\begin{center}
\includegraphics[width=0.9\linewidth,trim={1cm 0 0 0},clip]{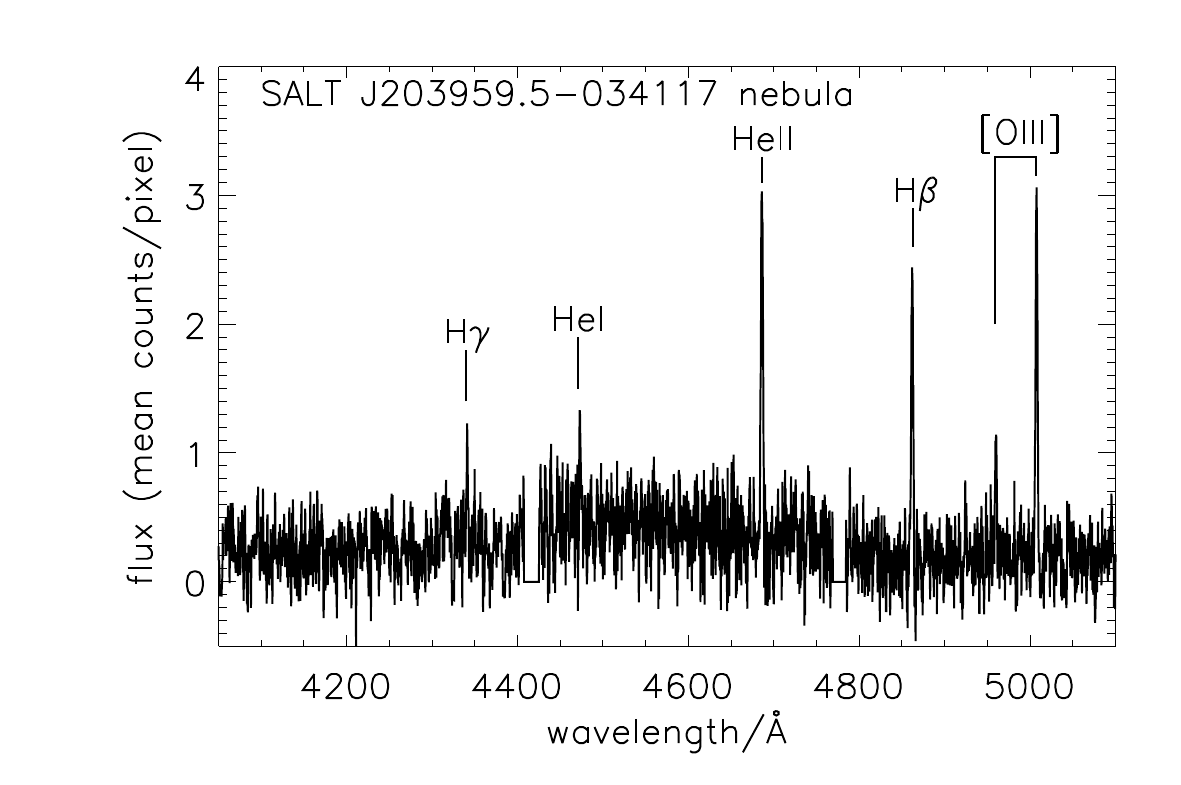}
\caption{Spectrum of nebula surrounding the O(H) star J2039, showing emission lines of H$\beta$, H$\gamma$, [O{\sc iii}], He{\sc i} and He{\sc ii}. Fluxes are shown as mean counts per pixel for a total integration time of 400\,s, and are extracted from a region  $1.9-18$" on both sides of the central star.}
\label{f_j2039rss}
\end{center}
\end{figure}

\begin{figure}
\begin{center}
\includegraphics[width=0.85\linewidth]{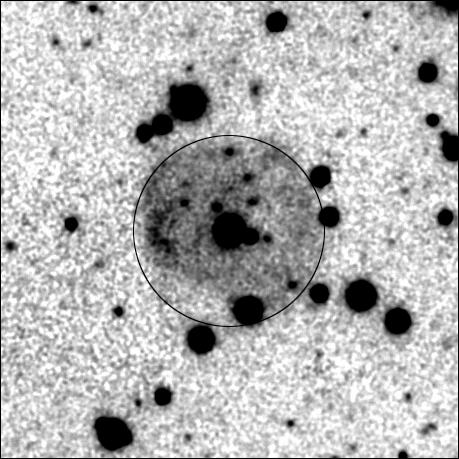}
\caption{A smoothed $g$-band image of O(H) star J2039 from the DECam legacy survey \citep{DECaLS19} revealing the surrounding planetary nebula. The circle has an angular diameter of 50" and the image measures $2\times2$ arcmin$^2$ with North up and East to the left.}
\label{f_j2039g}
\end{center}
\end{figure}

\section{Planetary Nebulae}
Since several O(H), O(He) and PG1159 stars are also the central stars of planetary nebulae, we checked the RSS long-slit spectra for evidence of H$\beta$ and [O{\sc iii}] emission lines. The O(H) star SALT J203959.5$-$034117 is the only one to show extended H$\beta$ emission, accompanied by O[{\sc iii}] emission at 5007 and 4959\AA, and He{\sc ii}4686\AA, with a total extent of some 38" (Fig.\,\ref{f_j2039rss}). H$\gamma$ and He{\sc i} 4471\AA\ are also visible in emission.  The PanSTARRS $g$-band and GALEX surveys show evidence of extended nebulosity, while the DECam Legacy Survey \citep{DECaLS19}
$g$-band shows a clear planetary nebula centred on J2039 with angular diameter 50" (Fig.\,\ref{f_j2039g}).  Being the central star of a PN, J2039 resembles the majority of O(H) stars. Following current practice, suggested names for the nebula include  PN G042.5$-$25.8 and  JeWeKi 1. SALT J212525.8$-$510559 is the hottest of the few O(H) stars without a detectable nebula.    


\begin{figure}
\begin{center}
\includegraphics[width=\linewidth]{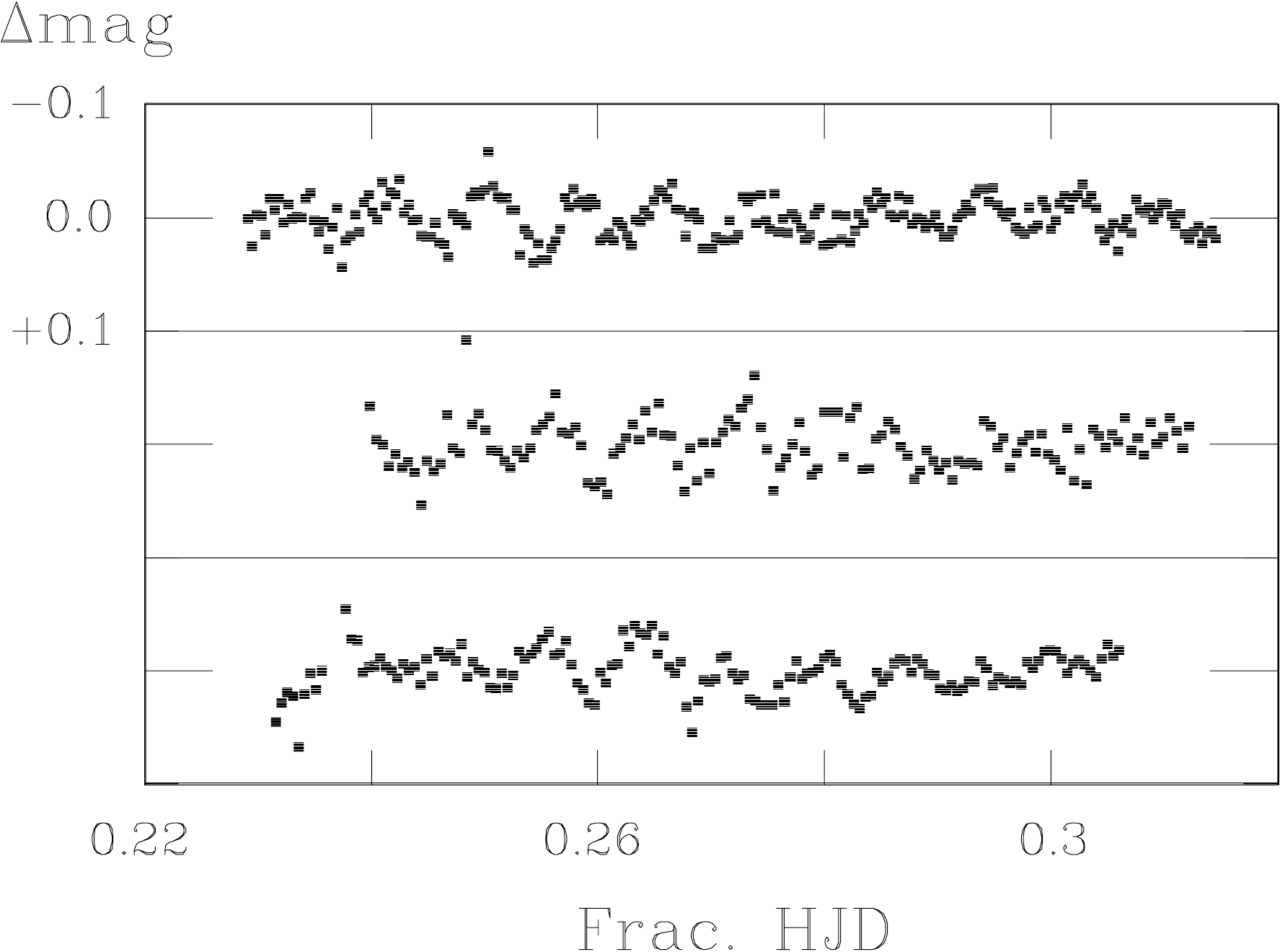}
\end{center}
\caption{Light curves obtained with the SAAO 1.0m telescope for the PG1159 star SALT J172411.7$-$632147. From the top, the data are from the nights 2022 Sep 14/15, 20/21 and 21/22. All data are differentially corrected and have the mean magnitude removed}
\label{f_phot1}
\end{figure}

\begin{figure}
\begin{center}
\includegraphics[width=\linewidth]{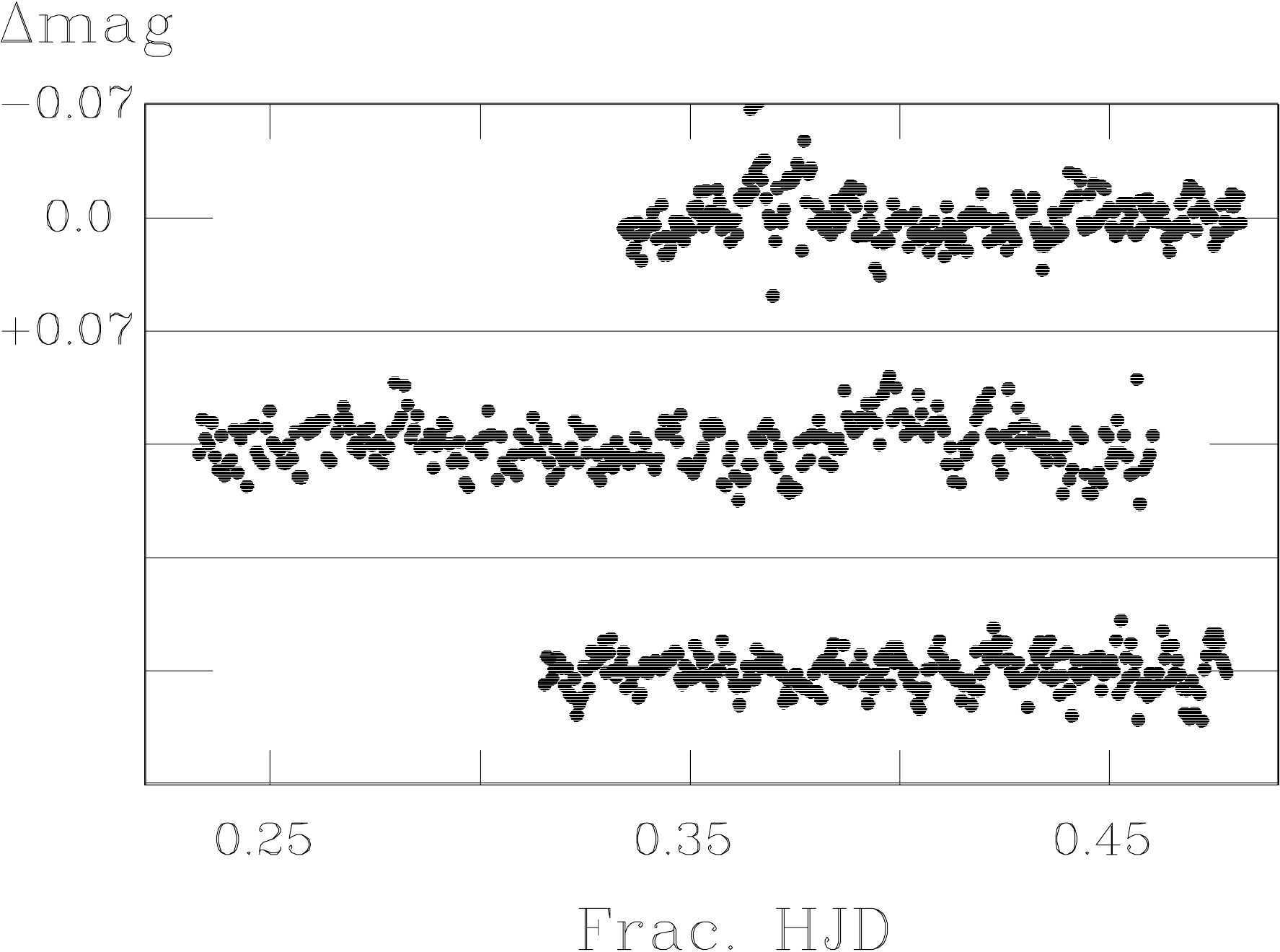}
\end{center}
\caption{Light curves obtained with the SAAO 1.0m telescope for PG1159 star SALT J213742.6$-$382901. From the top, the data are from the nights 2022 Sep 14/15, 19/20 and 21/22. All data are differentially corrected and have the mean magnitude removed.}
\label{f_phot2}
\end{figure}

\begin{figure}
\begin{center}
\includegraphics[width=\linewidth]{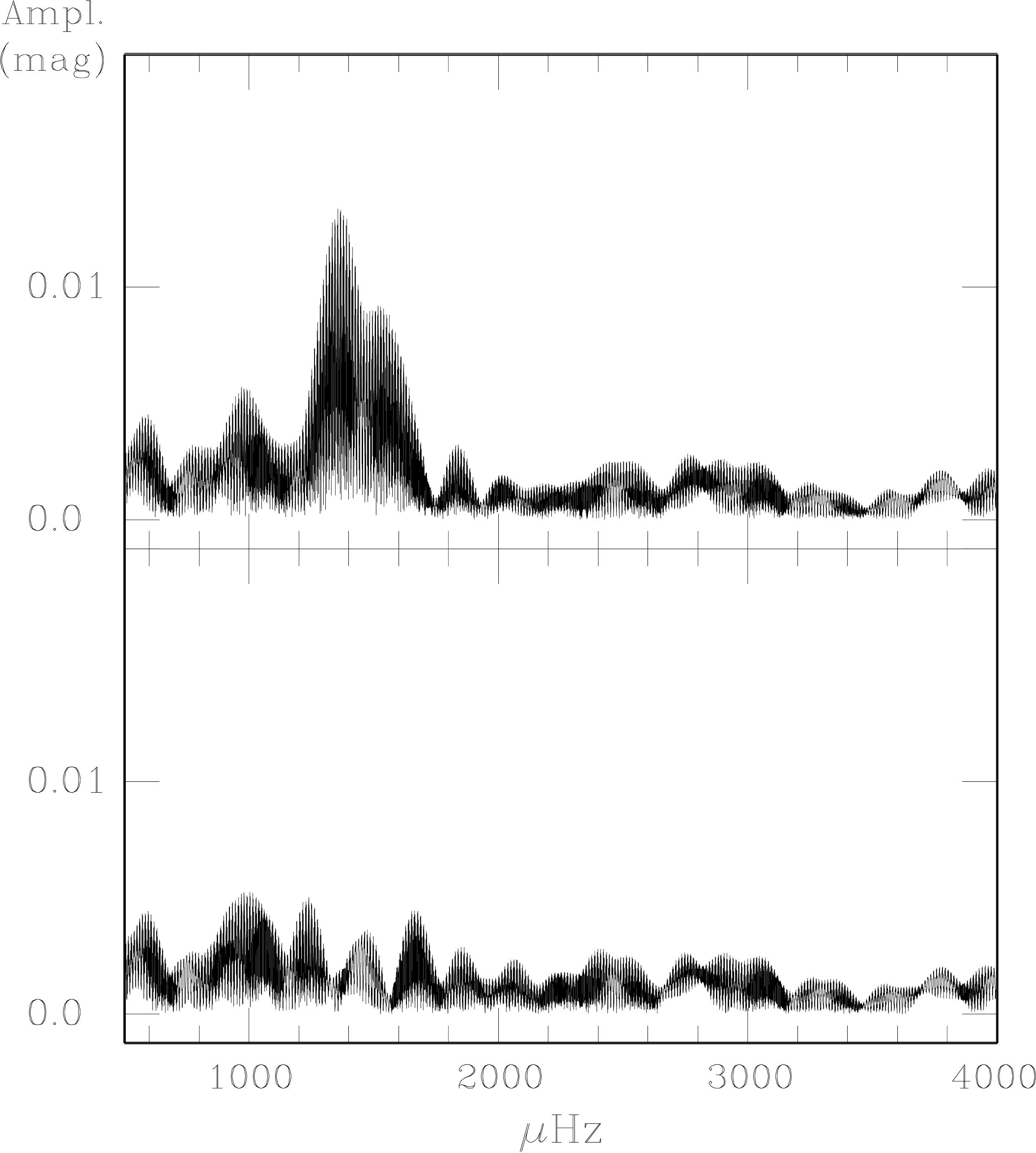}
\caption{Fourier transforms (FT) of the combined SAAO light curves of SALT J172411.7$-$632147. The upper panel shows the FT of the data shown in Fig.\,\ref{f_phot1}. The lower panel shows the FT of the same data pre-whitened by the first two frequencies shown for this star in Table\,\ref{t_freqs}.} 
\label{f_ft5}
\end{center}
\end{figure}

\begin{figure}
\begin{center}
\includegraphics[width=\linewidth]{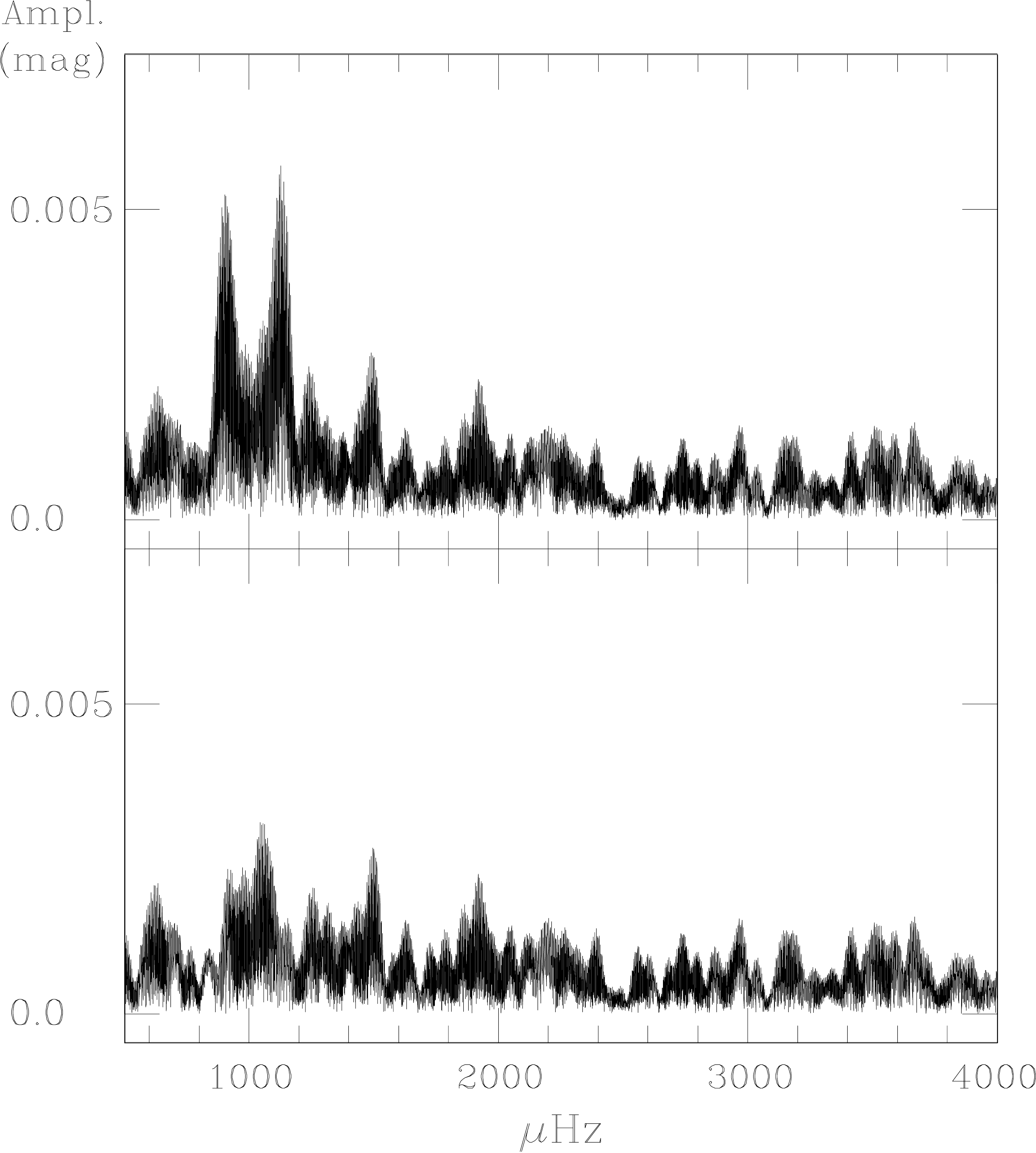}
\caption{Fourier transforms (FT) of the combined SAAO light curves of SALT J213742.6$-$382901. The upper panel shows the FT of the data shown in Fig.\,\ref{f_phot2}. The lower panel shows the FT of the same data pre-whitened by the first two frequencies shown for this star in Table\,\ref{t_freqs}}
\label{f_ft6}
\end{center}
\end{figure}

\begin{figure}
\begin{center}
\includegraphics[width=0.8\linewidth]{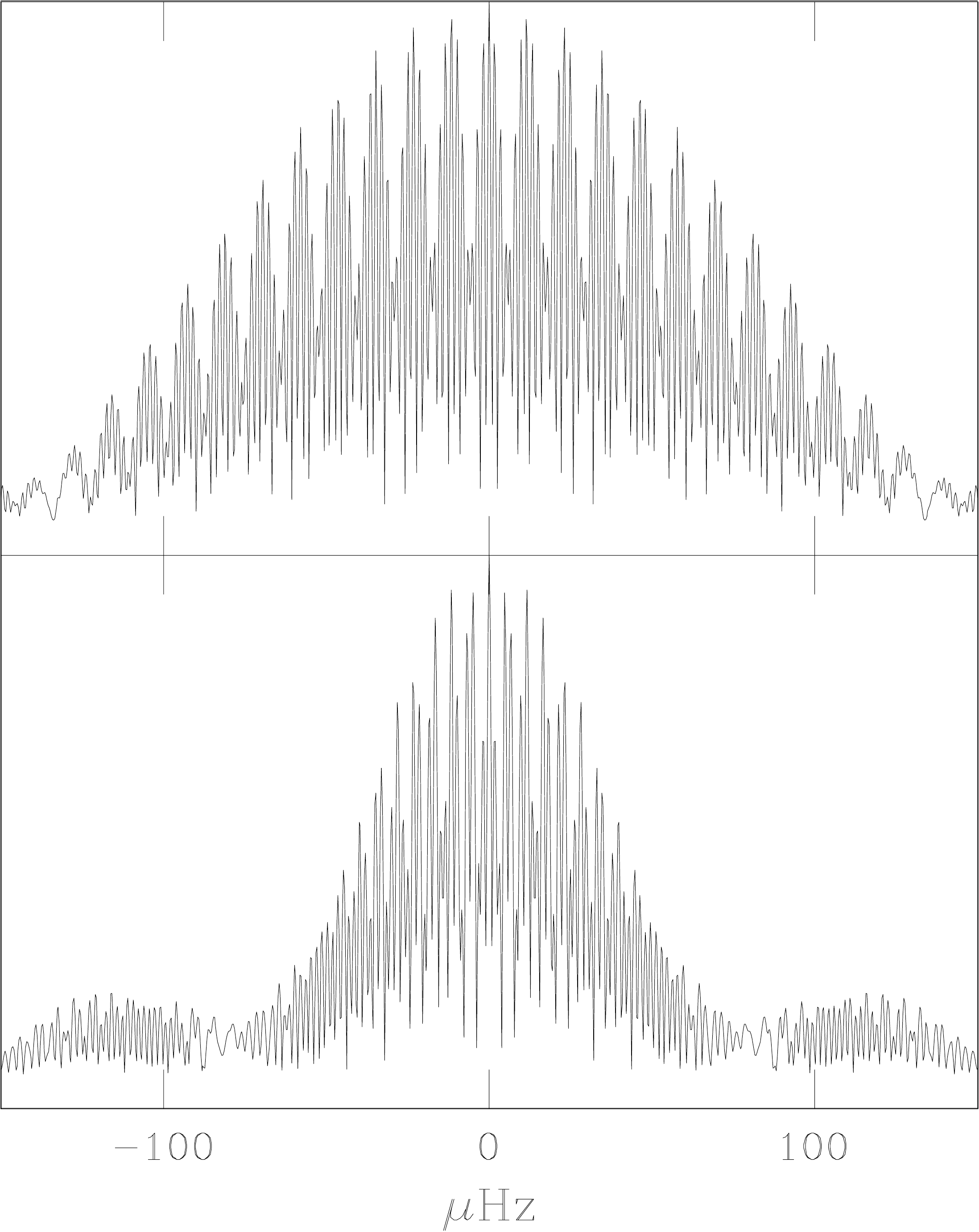}
\caption{Window functions for the  SAAO observations shown in Figs. \ref{f_phot1} (J1724: top) and \ref{f_phot2} (J2137: bottom). The scale is substantially expanded compared to Figs.\, \ref{f_ft5} and \ref{f_ft6} to show the alias pattern. For example, in the top panel the alias patterns caused by the 1 and 6 day separation of the data sets -- at 11.57 and $\sim$2 $\mu$Hz -- are clearly visible.}
\label{f_wfs}
\end{center}
\end{figure}

\begin{table}
\centering
\begin{tabular}{ccccc}
\hline
Date & JD & Exposure & n & Run\\
2022 Sep & 2450000+ & (sec) & (frames)& (hr) \\
 \hline
\multicolumn{3}{l}{SALT J172411.7$-$632147} \\[2mm] 
  14/15 &  9837 & 30  & 218 & 2.1 \\
  20/21 &  9843 & 45  & 129 & 1.7 \\
  21/22 &  9844 & 40  & 147 & 1.8 \\[2mm]
\hline 
\multicolumn{3}{l}{SALT J213742.6$-$382901} \\[2mm] 
  14/15 &  9837 & 40  & 279 & 3.4 \\
  19/20 &  9842 & 60  & 299 & 5.2 \\
  21/22 &  9844 & 45  & 278 & 3.8 \\[2mm]
\hline
\end{tabular}

\caption{Observing log for the SAAO SHOC data. In the SHOC mode used, there is a read-out time of $\sim$ 6 sec. for each frame.   }
\label{t_log}
\end{table}

\begin{table}
\centering
\begin{tabular}{ccc}
\hline
Frequency & Amplitude & Period \\
 ($\mu$Hz) & (mag) & (sec) \\
 \hline
\multicolumn{3}{l}{SALT J172411.7$-$632147} \\[2mm] 
  1356.62 (4) &   0.013 (1)  &     737  \\
  1528.05 (6) &   0.009 (1)  &     654  \\
  1238.9  (1) &   0.006 (1)  &     807  \\
  1003.3  (1) &   0.005 (1)  &     997  \\[2mm]
\hline 
\multicolumn{3}{l}{SALT J213742.6$-$382901} \\[2mm] 
  1127.14 (7) &  0.006 (1)  &  887  \\
   906.91 (6) &  0.005 (1)  & 1103  \\
  1045.2  (1) &  0.003 (1)  &  957  \\
  1500.8  (1) &  0.003 (1)  &  667  \\[2mm]
\hline
\end{tabular}
\caption{Frequencies, amplitudes and periods detected in the light curves of the two PG1159 stars in the sample. Numbers in brackets are the formal errors (s.d.) of the least squares solutions in the last digit of the quantity.  }
\label{t_freqs}
\end{table}

\section{SAAO Photometry}

Following the discovery that both PG1159 stars were of the N-rich variety, the authors postulated that they could be GW\,Vir variables, i.e., pulsating stars with multiple periods in the range 500 -- 1500 s. Fortunately, it proved possible to obtain data at very short notice using the SHOC photometer \citep[Sutherland High-speed Optical Camera][]{coppejans13} on the 1.0m Elizabeth Telescope at the South African Astronomical Observatory between 2022 September 14 and 21. Although bedevilled by bad weather and the drive motor shearing its mounting, it was possible to get three short runs on each star; a synopsis of the observing log is given in Table 3. No filter (``white light'') was used for all the observations in order to adequately resolve the anticipated periods likely to occur in these 16th magnitude stars. J1724 was already an hour past the meridian at the start of the night, but longer runs were possible on J2137; this proved fortuitous as the amplitudes of frequencies detected in the latter were much smaller than in the former and it was possible to show that both stars are variable on time scales consistent with GW\,Vir stars, as described below.

Reduction of the CCD frames and magnitude extraction were performed using software written by Darragh O'Donoghue and based on the {\sc DoPHOT} program described by \citet{schecter93}. For both stars it was possible to find several other, apparently non-variable, stars on the CCD frames which were used to differentially correct the target star data, allowing us to use data affected by small amounts of cloud. The resulting light curves for both stars are shown in Figs. \ref{f_phot1} and \ref{f_phot2}. 

The frequency analyses were carried out with Darragh O'Donoghue's {\sc EAGLE} program which uses the Fourier transform method of \citet{deeming75} as modified by \citet{kurtz85}. For each star, a periodogram was calculated for the light curve on each night, the highest peak removed by a least-squares fitted sinusoid \citep{deeming68} and the periodogram recalculated, and so on.
The few highest amplitude frequencies from each night were compared and four for each star were found to be reasonably coincident (occurring in two or three of the nights). The procedure was then repeated with all the data for each star and the same frequencies were recovered. These are illustrated in Figs. \ref{f_ft5} and \ref{f_ft6}. A  simultaneous fit to four frequencies for each star was then carried out and the results are listed in Table  \ref{t_freqs}. 

The two signals with highest amplitude for each star are well above their respective 4$\sigma$ detection thresholds ($\sim$0.001 for J1724 and $\sim$0.0007 for J2137) and we regard them as well-established. The two weaker signals in each star are close to those detection thresholds and thus less robust. A necessary caveat is illustrated by Fig. \ref{f_wfs}. The relatively short duration of each data set combined with the much longer time between them (1 and 6 days for J1724; 2 and 5 days for J2137) results in the possibility that the frequencies in Table \ref{t_freqs} might be aliases of the true frequencies. Nevertheless, it seems clear that both stars exhibit GW\,Vir variability.

Comparing the observed period ranges and effective temperatures with earlier observations and theoretical models of GW\,Vir variables \citep{corsico06,uzundag21}, both stars lie within expected ranges.
J2137 may be the hottest GW\,Vir found to date, but lies slightly to the blue (stable) side of the theoretical blue edge for non-radial $l=2$ g-modes, which is again blueward of the blue edge for $l=1$ modes \citep{corsico06}. 
Given the errors in spectroscopic $T_{\rm eff}$ and $\log g$, the discrepancy is not significant. 
However, higher quality spectroscopy of J2137 would help locate the position of the GW\,Vir $l=2$ stability boundary.


\section{Conclusion}
Serendipitously, the SALT survey for helium-rich subdwarfs has led to the discovery of eight hot white dwarfs and pre-white dwarfs with effective temperatures exceeding 100\,000\,K. They comprise two PG1159 stars, one DO white dwarf, three O(He) and two O(H) stars.  Follow-up spectroscopy and analysis will explore their significance for stellar evolution studies. One O(H) star is the central star of a newly-discovered planetary nebula. The other is the hottest 'naked' O(H) star. The DO white dwarf is also the hottest in its class. Both of the new PG1159 stars are variable with more than one period detected in the range 600 -- 1200s, giving them the characteristics of GW\,Vir variables. Again, one may be the hottest known GW\,Vir variable and an important test for theoretical models of stability in GW\,Vir stars. More extended observations, preferably from space, would allow their asteroseismic properties to be explored in more detail. 

\section*{Acknowledgments}
The observations reported in this paper were obtained with the Southern African Large Telescope (SALT) and with the Elizabeth Telescope of the South African Astronomical Observatory. 
CSJ and EJS are indebted to the UK Science and Technology Facilities Council via UKRI Grant Nos. ST/V000438/1 and ST/R504609/1,  and the Northern Ireland Department for Communities which funds the Armagh Observatory and Planetarium.  DK thanks the SAAO for continuing allocations of telescope time and the University of the Western Cape for funding to use those allocations.

This paper uses an image obtained with the Dark Energy Camera (DECam) as part of the Dark Energy Legacy Survey \citep{DECaLS19}.  
An extended acknowledgment for both projects can be found at {\tt https://www.legacysurvey.org/acknowledgment/}

For the purpose of open access, the author has applied a Creative Commons Attribution (CC BY) license to any Author Accepted Manuscript version arising.

\section*{Data Availability}

The SALT/RSS spectra used in this paper will be made public as part of the second data release for the SALT survey of helium-rich hot subdwarfs (Jeffery et al. in preparation). 
They will be made available following publication of this paper upon reasonable request to the authors.  
The photometric lightcurves will be made available on reasonable request to the authors. 
The model spectra will be made available on reasonable request to the authors. 

\bibliographystyle{mnras}
\bibliography{ehe}

\begin{thebibliography}{}
\makeatletter
\relax
\def\mn@urlcharsother{\let\do\@makeother \do\$\do\&\do\#\do\^\do\_\do\%\do\~}
\def\mn@doi{\begingroup\mn@urlcharsother \@ifnextchar [ {\mn@doi@}
  {\mn@doi@[]}}
\def\mn@doi@[#1]#2{\def\@tempa{#1}\ifx\@tempa\@empty \href
  {http://dx.doi.org/#2} {doi:#2}\else \href {http://dx.doi.org/#2} {#1}\fi
  \endgroup}
\def\mn@eprint#1#2{\mn@eprint@#1:#2::\@nil}
\def\mn@eprint@arXiv#1{\href {http://arxiv.org/abs/#1} {{\tt arXiv:#1}}}
\def\mn@eprint@dblp#1{\href {http://dblp.uni-trier.de/rec/bibtex/#1.xml}
  {dblp:#1}}
\def\mn@eprint@#1:#2:#3:#4\@nil{\def\@tempa {#1}\def\@tempb {#2}\def\@tempc
  {#3}\ifx \@tempc \@empty \let \@tempc \@tempb \let \@tempb \@tempa \fi \ifx
  \@tempb \@empty \def\@tempb {arXiv}\fi \@ifundefined
  {mn@eprint@\@tempb}{\@tempb:\@tempc}{\expandafter \expandafter \csname
  mn@eprint@\@tempb\endcsname \expandafter{\@tempc}}}

\bibitem[\protect\citeauthoryear{{Althaus}, {Panei}, {Miller Bertolami},
  {Garc{\'\i}a-Berro}, {C{\'o}rsico}, {Romero}, {Kepler}  \&
  {Rohrmann}}{{Althaus} et~al.}{2009}]{althaus09}
{Althaus} L.~G.,  {Panei} J.~A.,  {Miller Bertolami} M.~M.,
  {Garc{\'\i}a-Berro} E.,  {C{\'o}rsico} A.~H.,  {Romero} A.~D.,  {Kepler}
  S.~O.,   {Rohrmann} R.~D.,  2009, \mn@doi [\apj]
  {10.1088/0004-637X/704/2/1605}, \href
  {https://ui.adsabs.harvard.edu/abs/2009ApJ...704.1605A} {704, 1605}

\bibitem[\protect\citeauthoryear{{Bergeron} et~al.,}{{Bergeron}
  et~al.}{2011}]{bergeron11}
{Bergeron} P.,  et~al., 2011, \mn@doi [\apj] {10.1088/0004-637X/737/1/28},
  \href {https://ui.adsabs.harvard.edu/abs/2011ApJ...737...28B} {737, 28}

\bibitem[\protect\citeauthoryear{{Coppejans} et~al.,}{{Coppejans}
  et~al.}{2013}]{coppejans13}
{Coppejans} R.,  et~al., 2013, \mn@doi [\pasp] {10.1086/672156}, \href
  {https://ui.adsabs.harvard.edu/abs/2013PASP..125..976C} {125, 976}

\bibitem[\protect\citeauthoryear{{C{\'o}rsico}, {Althaus}  \& {Miller
  Bertolami}}{{C{\'o}rsico} et~al.}{2006}]{corsico06}
{C{\'o}rsico} A.~H.,  {Althaus} L.~G.,   {Miller Bertolami} M.~M.,  2006,
  \mn@doi [\aap] {10.1051/0004-6361:20065423}, \href
  {https://ui.adsabs.harvard.edu/abs/2006A&A...458..259C} {458, 259}

\bibitem[\protect\citeauthoryear{{Deeming}}{{Deeming}}{1968}]{deeming68}
{Deeming} T.~J.,  1968, \mn@doi [Vistas in Astronomy]
  {10.1016/0083-6656(68)90043-3}, \href
  {https://ui.adsabs.harvard.edu/abs/1968VA.....10..125D} {10, 125}

\bibitem[\protect\citeauthoryear{{Deeming}}{{Deeming}}{1975}]{deeming75}
{Deeming} T.~J.,  1975, \mn@doi [\apss] {10.1007/BF00681947}, \href
  {https://ui.adsabs.harvard.edu/abs/1975Ap&SS..36..137D} {36, 137}

\bibitem[\protect\citeauthoryear{{Dey} et~al.,}{{Dey} et~al.}{2019}]{DECaLS19}
{Dey} A.,  et~al., 2019, \mn@doi [\aj] {10.3847/1538-3881/ab089d}, \href
  {https://ui.adsabs.harvard.edu/abs/2019AJ....157..168D} {157, 168}

\bibitem[\protect\citeauthoryear{{Dreizler} \& {Werner}}{{Dreizler} \&
  {Werner}}{1996}]{dreizler96}
{Dreizler} S.,  {Werner} K.,  1996, \aap, \href
  {https://ui.adsabs.harvard.edu/abs/1996A&A...314..217D} {314, 217}

\bibitem[\protect\citeauthoryear{{Dreyer}}{{Dreyer}}{1888}]{dreyer88}
{Dreyer} J.~L.~E.,  1888, \memras, \href
  {https://ui.adsabs.harvard.edu/abs/1888MmRAS..49....1D} {49, 1}

\bibitem[\protect\citeauthoryear{{Drilling}, {Jeffery}, {Heber}, {Moehler}  \&
  {Napiwotzki}}{{Drilling} et~al.}{2013}]{drilling13}
{Drilling} J.~S.,  {Jeffery} C.~S.,  {Heber} U.,  {Moehler} S.,   {Napiwotzki}
  R.,  2013, \mn@doi [\aap] {10.1051/0004-6361/201219433}, \href
  {http://adsabs.harvard.edu/abs/2013A%26A...551A..31D} {551, A31}

\bibitem[\protect\citeauthoryear{{Dufour}, {Blouin}, {Coutu},
  {Fortin-Archambault}, {Thibeault}, {Bergeron}  \& {Fontaine}}{{Dufour}
  et~al.}{2017}]{dufour17}
{Dufour} P.,  {Blouin} S.,  {Coutu} S.,  {Fortin-Archambault} M.,  {Thibeault}
  C.,  {Bergeron} P.,   {Fontaine} G.,  2017, in {Tremblay} P.~E.,  {Gaensicke}
  B.,   {Marsh} T.,  eds,  Astronomical Society of the Pacific Conference
  Series Vol. 509, 20th European White Dwarf Workshop. p.~3 (\mn@eprint {arXiv}
  {1610.00986})

\bibitem[\protect\citeauthoryear{{Gaia Collaboration}}{{Gaia
  Collaboration}}{2021}]{gaia21.dr3}
{Gaia Collaboration} 2021, \mn@doi [\aap] {10.1051/0004-6361/202039657}, \href
  {https://ui.adsabs.harvard.edu/abs/2021A&A...649A...1G} {649, A1}

\bibitem[\protect\citeauthoryear{{Geier}, {Raddi}, {Gentile Fusillo}  \&
  {Marsh}}{{Geier} et~al.}{2019}]{geier19}
{Geier} S.,  {Raddi} R.,  {Gentile Fusillo} N.~P.,   {Marsh} T.~R.,  2019,
  \mn@doi [\aap] {10.1051/0004-6361/201834236}, \href
  {https://ui.adsabs.harvard.edu/abs/2019A&A...621A..38G} {621, A38}

\bibitem[\protect\citeauthoryear{{Gentile Fusillo} et~al.,}{{Gentile Fusillo}
  et~al.}{2019}]{gentilefusille19}
{Gentile Fusillo} N.~P.,  et~al., 2019, \mn@doi [\mnras]
  {10.1093/mnras/sty3016}, \href
  {https://ui.adsabs.harvard.edu/abs/2019MNRAS.482.4570G} {482, 4570}

\bibitem[\protect\citeauthoryear{{Jeffery}, {Miszalski}  \&
  {Snowdon}}{{Jeffery} et~al.}{2021}]{jeffery21a}
{Jeffery} C.~S.,  {Miszalski} B.,   {Snowdon} E.,  2021, \mn@doi [\mnras]
  {10.1093/mnras/staa3648}, \href
  {https://ui.adsabs.harvard.edu/abs/2021MNRAS.501..623J} {501, 623}

\bibitem[\protect\citeauthoryear{{Kilkenny}, {O'Donoghue}, {Koen}, {Stobie}  \&
  {Chen}}{{Kilkenny} et~al.}{1997}]{kilkenny97}
{Kilkenny} D.,  {O'Donoghue} D.,  {Koen} C.,  {Stobie} R.~S.,   {Chen} A.,
  1997, \mnras, \href {http://adsabs.harvard.edu/abs/1997MNRAS.287..867K} {287,
  867}

\bibitem[\protect\citeauthoryear{{Kilkenny}, {O'Donoghue}, {Worters}, {Koen},
  {Hambly}  \& {MacGillivray}}{{Kilkenny} et~al.}{2015}]{kilkenny15}
{Kilkenny} D.,  {O'Donoghue} D.,  {Worters} H.~L.,  {Koen} C.,  {Hambly} N.,
  {MacGillivray} H.,  2015, \mn@doi [\mnras] {10.1093/mnras/stv1771}, \href
  {http://adsabs.harvard.edu/abs/2015MNRAS.453.1879K} {453, 1879}

\bibitem[\protect\citeauthoryear{{Kilkenny}, {Worters}, {O'Donoghue}, {Koen},
  {Koen}, {Hambly}, {MacGillivray}  \& {Stobie}}{{Kilkenny}
  et~al.}{2016}]{kilkenny16}
{Kilkenny} D.,  {Worters} H.~L.,  {O'Donoghue} D.,  {Koen} C.,  {Koen} T.,
  {Hambly} N.,  {MacGillivray} H.,   {Stobie} R.~S.,  2016, \mn@doi [\mnras]
  {10.1093/mnras/stw916}, \href
  {http://adsabs.harvard.edu/abs/2016MNRAS.459.4343K} {459, 4343}

\bibitem[\protect\citeauthoryear{{Kurtz}}{{Kurtz}}{1985}]{kurtz85}
{Kurtz} D.~W.,  1985, \mn@doi [\mnras] {10.1093/mnras/213.4.773}, \href
  {https://ui.adsabs.harvard.edu/abs/1985MNRAS.213..773K} {213, 773}

\bibitem[\protect\citeauthoryear{{McGraw}, {Liebert}, {Starrfield}  \&
  {Green}}{{McGraw} et~al.}{1979}]{mcgraw79}
{McGraw} J.~T.,  {Liebert} J.,  {Starrfield} S.~G.,   {Green} R.,  1979, in
  {van Horn} H.~M.,  {Weidemann} V.,  eds, IAU Colloq. 53: White Dwarfs and
  Variable Degenerate Stars. pp 377--381

\bibitem[\protect\citeauthoryear{{Mendez}}{{Mendez}}{1991}]{mendez91}
{Mendez} R.~H.,  1991, in {Michaud} G.,  {Tutukov} A.~V.,  eds,  IAU Symposium
  Vol. 145, Evolution of Stars: the Photospheric Abundance Connection. p.~375

\bibitem[\protect\citeauthoryear{{Miller Bertolami}}{{Miller
  Bertolami}}{2016}]{millerbertolami16}
{Miller Bertolami} M.~M.,  2016, \mn@doi [\aap] {10.1051/0004-6361/201526577},
  \href {https://ui.adsabs.harvard.edu/abs/2016A&A...588A..25M} {588, A25}

\bibitem[\protect\citeauthoryear{{O'Donoghue}, {Kilkenny}, {Koen}, {Hambly},
  {MacGillivray}  \& {Stobie}}{{O'Donoghue} et~al.}{2013}]{odonoghue13}
{O'Donoghue} D.,  {Kilkenny} D.,  {Koen} C.,  {Hambly} N.,  {MacGillivray} H.,
   {Stobie} R.~S.,  2013, \mn@doi [\mnras] {10.1093/mnras/stt158}, \href
  {http://adsabs.harvard.edu/abs/2013MNRAS.431..240O} {431, 240}

\bibitem[\protect\citeauthoryear{{Reindl}, {Rauch}, {Werner}, {Kruk}  \&
  {Todt}}{{Reindl} et~al.}{2014}]{reindl14}
{Reindl} N.,  {Rauch} T.,  {Werner} K.,  {Kruk} J.~W.,   {Todt} H.,  2014,
  \mn@doi [\aap] {10.1051/0004-6361/201423498}, \href
  {https://ui.adsabs.harvard.edu/abs/2014A&A...566A.116R} {566, A116}

\bibitem[\protect\citeauthoryear{{Reindl}, {Geier}, {Kupfer}, {Bloemen},
  {Schaffenroth}, {Heber}, {Barlow}  \& {{\O}stensen}}{{Reindl}
  et~al.}{2016}]{reindl16}
{Reindl} N.,  {Geier} S.,  {Kupfer} T.,  {Bloemen} S.,  {Schaffenroth} V.,
  {Heber} U.,  {Barlow} B.~N.,   {{\O}stensen} R.~H.,  2016, \mn@doi [\aap]
  {10.1051/0004-6361/201527637}, \href
  {https://ui.adsabs.harvard.edu/abs/2016A&A...587A.101R} {587, A101}

\bibitem[\protect\citeauthoryear{{Schechter}, {Mateo}  \& {Saha}}{{Schechter}
  et~al.}{1993}]{schecter93}
{Schechter} P.~L.,  {Mateo} M.,   {Saha} A.,  1993, \mn@doi [\pasp]
  {10.1086/133316}, \href
  {https://ui.adsabs.harvard.edu/abs/1993PASP..105.1342S} {105, 1342}

\bibitem[\protect\citeauthoryear{{Stobie}, {Kilkenny}, {O'Donoghue}  \& {et
  al.}}{{Stobie} et~al.}{1997}]{stobie97a}
{Stobie} R.~S.,  {Kilkenny} D.,  {O'Donoghue} D.,   {et al.} 1997, \mnras,
  \href {http://adsabs.harvard.edu/abs/1997MNRAS.287..848S} {287, 848}

\bibitem[\protect\citeauthoryear{{Uzundag} et~al.,}{{Uzundag}
  et~al.}{2021}]{uzundag21}
{Uzundag} M.,  et~al., 2021, \mn@doi [\aap] {10.1051/0004-6361/202141253},
  \href {https://ui.adsabs.harvard.edu/abs/2021A&A...655A..27U} {655, A27}

\bibitem[\protect\citeauthoryear{{Werner} \& {Herwig}}{{Werner} \&
  {Herwig}}{2006}]{werner06}
{Werner} K.,  {Herwig} F.,  2006, \pasp, \href
  {http://ukads.nottingham.ac.uk/abs/2006PASP..118..183W} {118, 183}

\bibitem[\protect\citeauthoryear{{Werner} \& {Rauch}}{{Werner} \&
  {Rauch}}{2014}]{werner14}
{Werner} K.,  {Rauch} T.,  2014, \mn@doi [\aap] {10.1051/0004-6361/201424051},
  \href {https://ui.adsabs.harvard.edu/abs/2014A&A...569A..99W} {569, A99}

\bibitem[\protect\citeauthoryear{{Werner} \& {Rauch}}{{Werner} \&
  {Rauch}}{2015}]{werner15}
{Werner} K.,  {Rauch} T.,  2015, \mn@doi [\aap] {10.1051/0004-6361/201527261},
  \href {https://ui.adsabs.harvard.edu/abs/2015A&A...584A..19W} {584, A19}

\bibitem[\protect\citeauthoryear{{Werner}, {Rauch}  \& {Kruk}}{{Werner}
  et~al.}{2017}]{werner17}
{Werner} K.,  {Rauch} T.,   {Kruk} J.~W.,  2017, \mn@doi [\aap]
  {10.1051/0004-6361/201630266}, \href
  {https://ui.adsabs.harvard.edu/abs/2017A&A...601A...8W} {601, A8}

\bibitem[\protect\citeauthoryear{{Werner}, {Reindl}, {Dorsch}, {Geier},
  {Munari}  \& {Raddi}}{{Werner} et~al.}{2022}]{werner22}
{Werner} K.,  {Reindl} N.,  {Dorsch} M.,  {Geier} S.,  {Munari} U.,   {Raddi}
  R.,  2022, \mn@doi [\aap] {10.1051/0004-6361/202142397}, \href
  {https://ui.adsabs.harvard.edu/abs/2022A&A...658A..66W} {658, A66}

\bibitem[\protect\citeauthoryear{{Zhang} \& {Jeffery}}{{Zhang} \&
  {Jeffery}}{2012}]{zhang12a}
{Zhang} X.,  {Jeffery} C.~S.,  2012, \mn@doi [\mnras]
  {10.1111/j.1365-2966.2011.19711.x}, \href
  {http://adsabs.harvard.edu/abs/2012MNRAS.419..452Z} {419, 452}

\makeatother
\end{thebibliography}
\label{lastpage}

\end{document}